\documentclass[aps, twocolumn, showpacs]{revtex4}
\usepackage{amsfonts}
\usepackage{amssymb}
\usepackage{graphicx}
\usepackage{amsmath}
\usepackage[english]{babel}
\usepackage{color}

\begin{document}

\title{Symmetry and the critical phase of the two-bath spin-boson model: Ground-state properties}

\author{Nengji Zhou$^{1,2}$, Lipeng Chen$^{1}$, Dazhi Xu$^{1}$, Vladimir Chernyak$^{1,3}$, Yang Zhao$^{1}$\footnote{Electronic address:~\url{YZhao@ntu.edu.sg}}}
\date{\today}
\affiliation{$^1$Division of Materials Science, Nanyang Technological University, Singapore 639798, Singapore\\
$^2$Department of Physics, Hangzhou Normal University, Hangzhou 310046, China\\
$^3$Department of Chemistry, Wayne State University, Detroit, USA}

\begin{abstract}
A generalized trial wave function termed as the ``multi-${\rm D}_1$ ansatz'' has been developed to study the ground state of the spin-boson model with simultaneous diagonal and off-diagonal coupling in the sub-Ohmic regime. Ground state properties including the energy and the spin polarization are investigated, and the results are consistent with those from the exact diagonalization and density matrix renormalization group approaches for the cases involving two oscillators and two baths described by a continuous spectral density function. Breakdown of the rotational and parity symmetries along the continuous quantum phase transition separating the localized phase from the critical phase has been uncovered. Moreover, the phase boundary is determined accurately with the corresponding symmetry parameters of the rotational and parity symmetries. A critical value of the spectral exponent $s^* = 0.49(1)$ is predicted in the weak coupling limit, which is in agreement with the mean-field prediction of $1/2$, but much smaller than the earlier literature estimate of $0.75(1)$.
\end{abstract}

\maketitle

\section{Introduction}

The paradigm of a quantum spin interacting with its dissipative environment has drawn sustained research interests in a variety of fields including quantum computation \cite{qs1,qs2,qs3}, spin dynamics \cite{Leggett,dyna,dua}, quantum phase transitions \cite{qpt1,qpt2,alv,win}, charge transfer in biological molecules \cite{et1,et2} and impurity effects in magnetic materials \cite{wilson, stamp, vajk}. Among the most popular models employed in this regards is the spin-boson model \cite{Weiss} that describes a two-level system, i.e., a spin $1/2$, coupled linearly to an environment represented by a set of harmonic oscillators. The coupling between the system and the environment can be characterized by a spectral function $J(\omega)$. This model is known to exhibit rich ground state properties. In particular, if the bath is characterized by a gapless spectral density $J(\omega) \sim 2\alpha \omega^s$, a quantum phase transition is expected to appear, separating a non-degenerate ``delocalized'' phase from a doubly degenerate ``localized'' phase due to the competition between tunneling and environment-induced dissipation. Depending on the value of $s$, there exist three distinct cases known as sub-Ohmic ($s<1$), Ohmic ($s=1$) and super-Ohmic ($s>1$) regimes. Recent theoretical studies \cite{qpt1,qpt2,alv,win} show that the transition is of second order in the sub-Ohmic regime and Kosterlitz-Thouless type in the Ohmic regime \cite{Leggett}. In the super-Ohmic regime, however, there is no phase transition.

\begin{figure}[tbp]

\centering
\includegraphics*[width=0.7\linewidth]{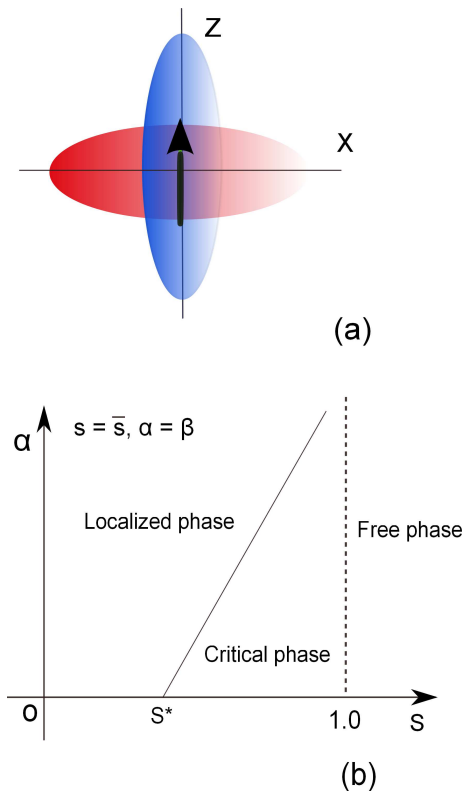}
\caption[FIG]{(Color online) (a) Schematic of the two-bath spin-boson model. A single spin is immersed in two independent baths with simultaneous diagonal coupling (Z) and off-diagonal coupling (X). (b) Schematic plot of the phase diagram of the spin-boson model with two identical bosonic baths $s=\bar{s}$ and $\alpha=\beta$.
Where $s$ ($\bar{s}$) and $\alpha$ ($\beta$) represent the spectral exponents and coupling strengths, respectively, for the spectral density functions $J_z(\omega)$ ($J_x(\omega)$). Three different phases (localized, critical and free) are displayed in the $\alpha$-$s$ plane with two critical values of the spectral exponents, $s^{*}$ and $1.0$. }
\label{f0}
\end{figure}

A number of studies have investigated  extensions of the standard spin-boson model, for example, to a two-spin system involving a common bath \cite{orth, mcc} or two independent baths \cite{wan13}, and to a single spin coupled to a bath with simultaneous diagonal and off-diagonal coupling \cite{lv}. We have recent studied a two-bath spin-boson model \cite{zhao14, zhou14}, shown schematically in Fig.~\ref{f0}(a)], where $\rm X$ and $\rm Z$ denote the diagonal and off-diagonal coupling, respectively, and the arrow represents a spin. The bath spectral densities can be described by $J_z(\omega)  =  2\alpha\omega_c^{1-s}\omega^s, ~J_x(\omega)  =  2\beta\omega_c^{1-\bar{s}}\omega^{\bar{s}}$, where $\alpha$ and $\beta$ are the dimensionless coupling strengths, and $s$ and $\bar{s}$ denote the spectral exponents characterizing the two baths coupled to the spin diagonally and off-diagonally, respectively. Possible realizations of such two-bath model include impurities in a magnet coupled to two spin-wave modes or two sources of dissipation \cite{kot98,voj00,net03, khv04}, excitonic energy transfer processes in natural and artificial light-harvesting systems \cite{pach}, electromagnetic fluctuations of two linear circuits attached to a superconducting qubit \cite{you, card,raft}, two cavity fields coupled to a SQUID-based charge qubit \cite{liao}, and the process of thermal transport between two reservoirs coupled with a molecular junction \cite{ruok}.

In the two-bath model with $s=\bar{s},~\alpha=\beta$, studies based on the perturbative renormalization group theory predict the presence of two phases, namely,  the ``critical phase'' and the ``free phase," in the absence of bias and tunneling \cite{voj00,Sen00, Zhu02}. Very recently, the existence of a ``localized phase" in the two-bath model was discovered numerically in the strong coupling regime \cite{guo}. The schematic of the phase diagram that emerges is shown in Fig.~\ref{f0}(b). A continuous quantum phase transition separating the localized phase from the critical phase was claimed to exist only for the spectral exponent $s^{*} < s <1$, and a critical value of the spectral exponent, $s^{*}=0.75(1)$, was estimated from the density matrix renormalization group (DMRG) calculations.  When $s>1$, the impurity behaves as a free spin in the so-called free phase \cite{guo}. The phase boundary was determined from the response to the external field (i.e., the bias or tunneling) perpendicular to the bath plane. However, the localized-to-delocalized phase transition will occur under the external field, which renders the phase diagram very complicated. Still unclear is the influence of the external field to the localized-to-critical transition. Moreover, the critical value of the spectral exponent was predicted by a recent mean-field analysis \cite{zhou14} to $s^{*}=1/2$, that stands at variance to the aforementioned DMRG result. It thus remains a challenging task to map out precisely the localized-to-critical phase transition as represented in Fig.~\ref{f0}(b) in the absence of an external field.

In the absence of bias and tunneling, the two-bath model exhibits a high level of symmetry, including the parity symmetry and rotational symmetry \cite{zhao14, zhou14, bru14}. In the localized phase, spontaneous symmetry breaking takes place due to strong spin-bath coupling. Hence, a symmetry analysis may help distinguish the critical and localized phases. In addition, a novel quantum phase transition from a doubly degenerate ``localized phase'' to another doubly degenerate ``delocalized phase'' is uncovered with respect to the ratio of the coupling strengths $\alpha/\beta$ within the two baths \cite{zhou14}. The transition is inferred to be of the first order, and the transition point $\alpha/\beta=1$ is determined when the spectral exponents of the two baths are identical. Since the system at the transition point corresponds to the $\rm XZ$-symmetric spin-boson model, the critical properties of the ground state, i.e., the spin polarization $m(\alpha,s)$ and generalized susceptibility $\chi(\alpha, s)$ at $s=\bar{s}$ and $\alpha=\beta$, can also be used to distinguish the localized and critical phases.

The purpose of this paper is to investigate various ground-state phases in the extended spin-boson model involving two identical, independent baths, and determine the critical value of the spectral exponent $s^{*}$ separating the localized and critical phases in the weak coupling limit of $\alpha \rightarrow 0$. Via the variational approach, the DMRG approach, and the exact diagonalization method, we conduct a comprehensive study on the ground state properties of the two-bath spin-boson model with zero bias and tunneling for the baths described by single mode as well as continuous spectral density function. In this work, rotational and parity symmetry breaking is found to occur along the localized-to-critical phase transition, and the phase boundary is obtained with $s^{*}=0.49(1)$, consistent with the mean-field predictions.

The rest of the paper is organized as follows. In Sec.~II, the two-bath spin-boson model and its symmetry properties are described, along with an introduction to the variational approach.  Sec.~III and IV present the numerical results for the localized and critical phases in the two-bath spin-boson model involving a spin coupled to two oscillators or two baths described by a continuous spectral density function, respectively. Conclusions are drawn in the final Sec.~V.

\section{Methodology}

\subsection{Model}

The two-bath spin-boson model can be described by the Hamiltonian below
\begin{eqnarray}
\hat{H} & = & \frac{\varepsilon}{2}\sigma_z-\frac{\Delta}{2}\sigma_x + \sum_{l,i} \omega_l b_{l,i}^\dag b_{l,i} \nonumber \\
 & + & \frac{\sigma_z}{2}\sum_l \lambda_l(b^\dag_{l,1}+b_{l,1})  \nonumber \\
 & + & \frac{\sigma_x}{2}\sum_l \phi_l(b^\dag_{l,2}+b_{l,2}),
\label{Ohami}
\end{eqnarray}
where $\varepsilon$ and $\Delta$ is the spin bias and tunneling constant, respectively, $i=1,2$ is the index of the baths, and $\lambda_l$ ($\phi_l$) is the diagonal (off-diagonal) coupling strength.  In order to investigate the quantum phase transition between the critical and localized phases, we focus on the case of $\varepsilon = \Delta = 0$ as mentioned in the `Introduction'.  A logarithmic discretization procedure is adopted by dividing the phonon frequency domain $[0, \omega_c]$ into $M$ intervals $\omega_c[\Lambda^{-l},\Lambda^{-(l-1)}]$ ($l=1,2,\ldots, M)$ \cite{Bulla, zyy}. The coupling strengthes $\omega_l$  and $\lambda_l$ (or $\phi_l$) in Eq.~(\ref{Ohami}) can then be calculated as
\begin{eqnarray}
\lambda_l^2 & = & \int^{\Lambda^{-l}\omega_c}_{\Lambda^{-l-1}\omega_c}dt J(t), \nonumber \\
\omega_l & = & \lambda^{-2}_l \int^{\Lambda^{-l}\omega_c}_{\Lambda^{-l-1}\omega_c}dtJ(t)t,
\label{sbm1_dis}
\end{eqnarray}
For convenience, the frequency cut off $\omega_c=1$ and the discretization factor $\Lambda=2$ are set throughout this paper. It should be noted that infinite bath modes are considered via the integration of the continuous spectral density $J(\omega)$, although the number of effective bath modes $M$ is finite.

Since various values of $\langle \sigma_x\rangle$ and $\langle\sigma_{z}\rangle$ are possible due to the extended symmetry of the two-bath model with zero bias and tunneling, the spin polarization is introduced as
\begin{equation}
m = \sqrt{\langle \sigma_x \rangle^2 + \langle \sigma_y \rangle^2 + \langle \sigma_z \rangle^2}
\label{spin polarization}
\end{equation}
Due to Hamiltonian invariance under the transformation $\sigma_y\rightarrow{-\sigma_y}$, the $y$ component of the spin polarization is $\langle{\sigma_y}\rangle \equiv 0$.
Hence, $m$ can be simplified to be $\sqrt{\langle{\sigma_x}\rangle^2+\langle{\sigma_z}\rangle^2}$. According to Ref.\cite{guo}, the
critical phase is characterized by $\langle{\sigma_i}\rangle=0$ $(i=x,y,z)$, which results in $m=0$.

In addition, the phonon population $P_{\rm ph}(x,z)$ is introduced to depict the ground state of the two-bath model.
Assuming that the wave function of the ground state can be written as
\begin{equation}
|\Psi_{g}\rangle =  | + \rangle |\psi_{+}\rangle_{\rm ph} +  |-\rangle |\psi_{-}\rangle_{\rm ph},
\label{gs}
\end{equation}
where $|\psi_{+}\rangle_{\rm ph}$ and $|\psi_{-}\rangle_{\rm ph}$ are the phonon parts of the wave function corresponding to the spin up and down states, respectively, which can be expanded in a series of Fock states or coherent states. The phonon population $P_{\rm ph}(x,z)$ is thus defined as
\begin{eqnarray}
P_{\rm ph}(x,z) & = & \langle \Psi_g(x,z)|\Psi_g(x,z)\rangle_{\rm spin} \nonumber  \\
& = &   |\psi_{+}(x,z)|^2 + |\psi_{-}(x,z)|^2
\label{density}
\end{eqnarray}
where $\langle \cdots \rangle_{\rm spin}$ represents the trace of the spin freedom in the wave function, $x$ and $z$ are coordinates in $\rm X-Z$ plane corresponding to the off-diagonal and diagonal coupling baths, and $\psi_{\pm}(x,z)=\langle \vec{r} |\psi_{\pm}\rangle_{\rm ph}$ is the phonon-component of wave function in the two-dimensional coordinate representation
$\vec{r} = (x, z)$.

\subsection{Ground state symmetry}

\begin{figure}[tbp]
\centering
\begin{minipage}[b]{0.5\textwidth}
\includegraphics*[width=0.5\linewidth]{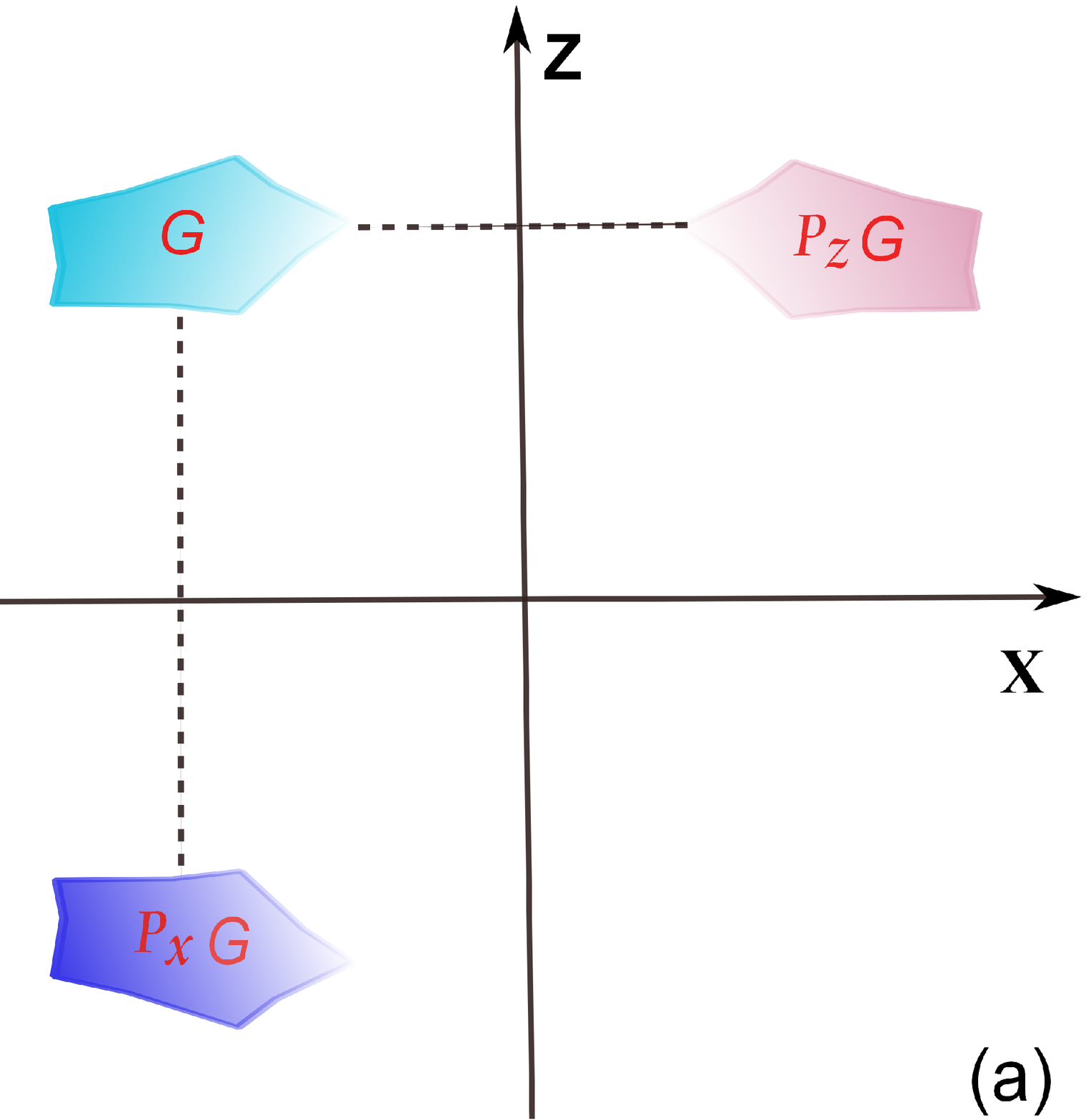} \\
\vspace{2\baselineskip}
\includegraphics*[width=0.5\linewidth]{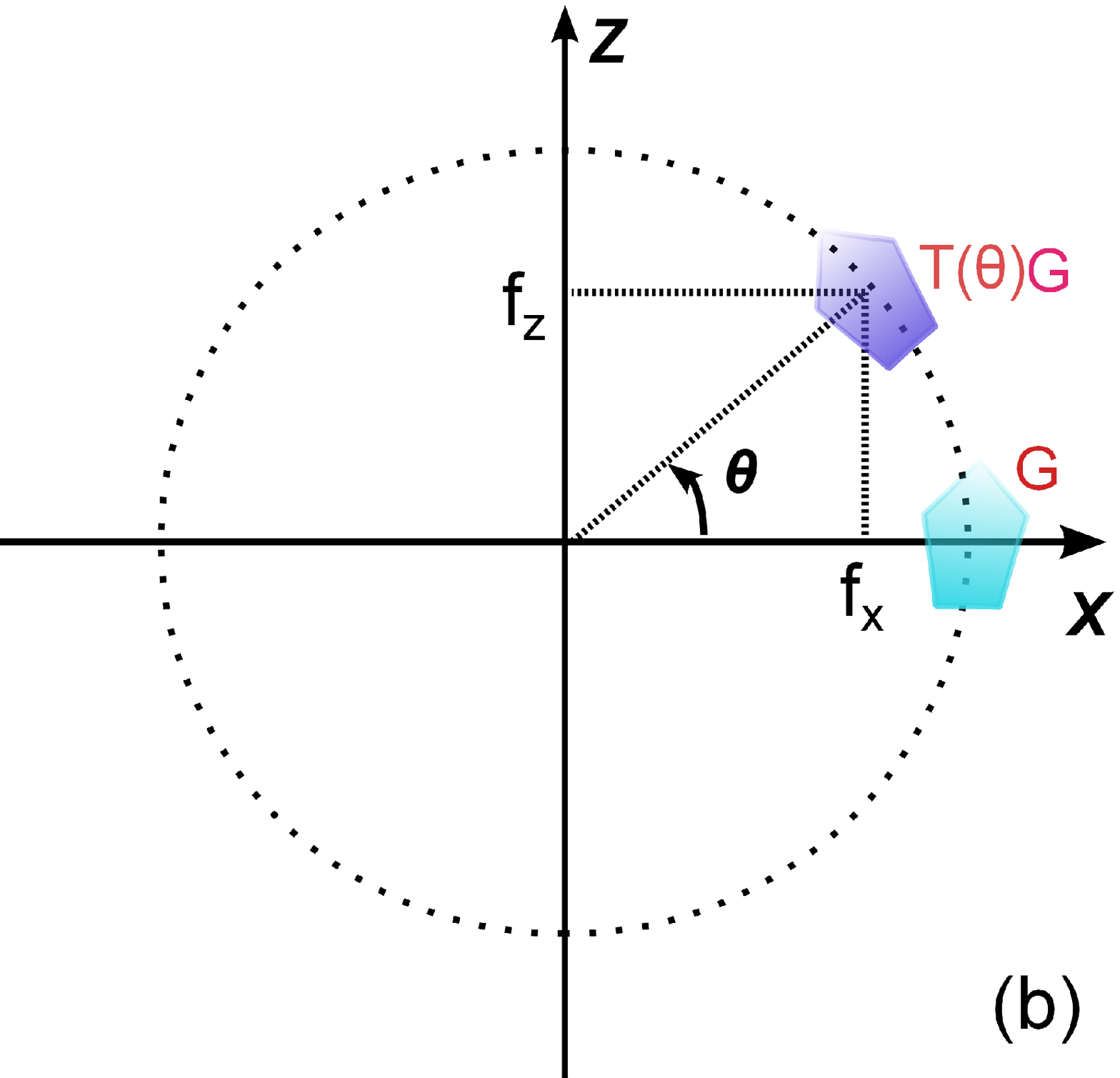}
\end{minipage}
\caption[FIG]{(Color on-line) The schematic of the influence of the parity symmetry operators ${\cal P}_{x},{\cal P}_{z}$ and rotational symmetry operator $\hat{T}(\theta)$ to the ground state $G$ are displayed in (a) and (b), respectively. For convenience, we use the polygon shape of the ground states to emphasize the influence of symmetry operations.
${\cal P}_x G, {\cal P}_z G$ and $\hat{T}(\theta) G$ are new ground states under these symmetry operations, and $f_x, f_z$ represent the coordinate values of the ground state in the $\rm X$ and $\rm Z$ directions, respectively.
}
\label{symmetry}
\end{figure}

The model studied in this paper exhibits a high level of symmetry due to zero bias and tunneling ($\varepsilon=0, \Delta=0$). A group theory analysis \cite{zhou14, zhao14} shows that the ground state is always doubly degenerate. We introduce four parity symmetry operators including ${\cal I}={\rm id}$ as the unit operator,
\begin{eqnarray}
\label{symmetry-generators} \mathcal{P}_{x} & = & \sigma_{x}  \textrm{e}^{i\pi \sum_{l}b^{\dagger}_{l,1}b_{l,1}}, \\ \nonumber
\mathcal{P}_{z} & = & \sigma_{z} \textrm{e}^{i\pi\sum_{l}b^{\dagger}_{l,2}b_{l,2}},
\end{eqnarray}
and ${\cal P}_x{\cal P}_z$. The influence of the parity symmetry operations to the ground state $G$ is displayed in Fig.~\ref{symmetry}(a). Under the operation ${\cal P}_{x}$ (${\cal P}_{z}$ ), the sign of the coordinate values corresponding to the displacements of phonons in the diagonal coupling bath (off-diagonal coupling bath) will be changed.
The symmetry parameters $\zeta_x$ and $\zeta_z$ of the parity symmetry are defined as
\begin{eqnarray}
\zeta_x & = & \langle \Psi_g|{\cal P}_x|\Psi_g\rangle, \nonumber \\
\zeta_z & = & \langle \Psi_g|{\cal P}_z|\Psi_g\rangle.
\label{order_parameter}
\end{eqnarray}
When $\alpha \neq \beta$, the results $\zeta_z=1$ and $\zeta_x=0$ ($\zeta_z=0$ and $\zeta_x=1$) are obtained for the localized phase (delocalized phase) in the two-bath spin boson \cite{zhou14}. However, $\zeta_x$ and $\zeta_z$ in the case of $\alpha=\beta$ are still unclear. A vanishing value of $\zeta=\sqrt{\zeta_x^2+\zeta_z^2}$ usually indicates breakdown of the parity symmetry. In contrast, one has $\zeta = 1$ when the ground state has perfect parity symmetry along the X or Z direction. If $0<\zeta<1$, the ground state exhibits partial parity symmetry which may be induced by the numerical errors or the finite number of the degrees of freedom.

In the $\rm XZ$-symmetric spin-boson model with $s=\bar{s}$ and $\alpha=\beta$, the system may exhibit a rotational symmetry, since the Hamiltonian is invariant when one simultaneously rotates the spin and the two baths in the $\rm X$-$\rm Z$ plane by an arbitrary angle $\theta$. According to the Abelian $U(1)$ symmetry of the two-bath model proposed in Ref.~\cite{bru14}, the rotational
symmetry operator $\hat{T}(\theta)$ is introduced as
\begin{equation}
\hat{T}(\theta)=\exp(-i\theta\hat{S}), \label{rotation}
\end{equation}
where $\hat{S}$ is the generator of the $U(1)$ symmetry defined as
\begin{equation}
\hat{S}=\frac{1}{2}\sigma_y + i\sum_{l=1}^M\left(b_{l,1}b_{l,2}^{\dag}-b_{l,1}^{\dag}b_{l,2}\right).
\label{generator}
\end{equation}
Figure.~\ref{symmetry}(b) shows the influence of the rotational symmetry operator on the ground state $G$, where the coordinate values of the ground state $f_x, f_z$ in the $\rm X$ and $\rm Z$ directions are proportional to the displacement coefficients of the phonons in the off-diagonal coupling and diagonal coupling baths, respectively.

The symmetry parameters $\gamma(\theta)$ and $\gamma_{\rm ph}(\theta)$ are introduced to quantitatively measure the rotational symmetry as
\begin{eqnarray} \label{order_rotation}
\gamma(\theta) & = & \langle \Psi_g|\hat{T}(\theta)|\Psi_g\rangle, \\ \nonumber
\gamma_{\rm ph}(\theta) &=& \langle \Psi_g|\hat{T}_{\rm ph}(\theta)|\Psi_g\rangle \\ \nonumber
 & = & \langle \Psi_g|\exp\left[\sum_{l=1}^M\theta\left(b_{l,1}b_{l,2}^{\dag}-b_{l,1}^{\dag}b_{l,1}\right)\right]|\Psi_g\rangle.
\end{eqnarray}
It should be noted that only the rotation of the two baths is considered in the definition of $\gamma_{\rm ph}(\theta)$.
The system energy $E(\theta)=\langle\Psi_g|\hat{T}^{\dag}(\theta)\hat{H}\hat{T}(\theta)|\Psi_g\rangle$ is expected to be independent of the rotational angle $\theta$ in the two-bath spin-boson model involving two identical baths. If the ground state has perfect rotational symmetry, one has $\gamma_{\rm ph}(\theta)=1$ for the whole regime of rotational angle $\theta$. In contrast, $\gamma_{\rm ph}(\theta) \sim \delta(\theta)$ is obtained in the absent of the rotational symmetry, where $\delta(\theta)$ is a delta function. When the ground state has partial rotational symmetry, $\gamma_{\rm ph}(\theta)$ decays with the rotational angle $\theta$.
Furthermore, we investigate the $\theta$-dependent behavior of the parity-symmetry parameters $\zeta(\theta)$ defined as
\begin{eqnarray}
\zeta_x(\theta) & = & \langle\Psi_g|\hat{T}_{\rm ph}^\dag(\theta){\cal P}_{x} \hat{T}_{\rm ph}(\theta)|\Psi_g\rangle   \nonumber \\
\zeta_z(\theta) & = &\langle\Psi_g|\hat{T}_{\rm ph}^\dag(\theta){\cal P}_{z} \hat{T}_{\rm ph}(\theta)|\Psi_g\rangle    \nonumber \\
\zeta(\theta) & = & \sqrt{\zeta_x(\theta)^2 + \zeta_z(\theta)^2},
\label{zeta_angle}
\end{eqnarray}
where $\hat{T}_{\rm ph}(\theta)$ defined in Eq.~(\ref{order_rotation}) is the phonon part of the rotational operator, and $\hat{T}_{\rm ph}(\theta)|\Psi_g\rangle$ is
one of the degenerate ground states obtained by rotating the ground state $|\Psi_g\rangle$. In the rest of the paper, both symmetry parameters, $\gamma_{\rm ph}(\theta)$ and $\zeta(\theta)$ of the rotational and parity symmetries, respectively, will be comprehensively studied, as they are useful and sensitive to detect the spontaneous symmetry breaking in the localized-to-critical phase transition.

\subsection{Variational method}

A systematic coherent-state expansion of the ground state wave function, termed as the ``multi-${\rm D}_1$ ansatz,'' is introduced as the variational trial ansatz \cite{zhou14,Bera, Bera_2}. It can be written as
\begin{eqnarray}
|\Psi \rangle & = & |+\rangle \sum_{n=1}^{N} A_n \exp\left[ \sum_{l}^{2M}\left(f_{n,l}b_l^{\dag} - \mbox{H}.\mbox{c}.\right)\right] |0\rangle_{\textrm{ph}} \nonumber \\
              & + & |-\rangle \sum_{n=1}^{N} B_n \exp\left[ \sum_{l}^{2M}\left(g_{n,l}b_l^{\dag} - \mbox{H}.\mbox{c}.\right)\right] |0\rangle_{\textrm{ph}},
\label{vmwave}
\end{eqnarray}
where H$.$c$.$ denotes Hermitian conjugate, $|+\rangle$ ($|-\rangle$) stands for the spin up (down) state, $|0\rangle_{\textrm{ph}}$
is the vacuum state of the phonon bath, and $M$ and $N$ represent the numbers of the bath modes and coherent superposition states, respectively.
In fact, Eq.~(\ref{vmwave}) describes a superposition of the spin states $|\pm\rangle$ that are correlated
with the bath modes with displacements $f_{n,l}$ and $g_{n,l}$,
where $n$ and $l$ represent the ranks of the coherent superposition state and effective bath mode, respectively.
The displacements ($f_{n,l}, g_{n,l}$) with $0<l\leq M$ ($M<l\leq 2M$) correspond to the phonons in the diagonal (off-diagonal) coupling bath.
Using this trial wave function, the system energy $E$ can be calculated with
the Hamiltonian expectation $H=\langle\Psi|\hat{H}|\Psi\rangle$ and the norm of the wave function
$D=\langle\Psi|\Psi\rangle$ as $E=H/D$. The ground state is then obtained by minimizing
the energy with respect to the variational parameters $A_n, B_n, f_{n,l}$ and $g_{n,l}$.
The variational procedure entails $N(4M+2)$ self-consistent equations,
\begin{equation}
\frac{\partial H}{\partial x_{i}} - E\frac{\partial D}{\partial x_{i}} = 0,
\label{vmit}
\end{equation}
where $x_i(i=1,2,\cdots, 4NM+2N)$ denote the variational parameters. The ``multi-${\rm D}_1$'' ansatz is much more sophisticated and contains more
flexible variational parameters than the Silbey-Harris ansatz \cite{sil84} and Nazir's ansatz \cite{naz12}, where only $2M+1$ and $4M+2$ variational parameters
are employed, respectively. For example, if $N=16$ and $M=20$, our new ansatz has $1312$ variational parameters, compared to $41$ parameters in the Silbey-Harris ansatz  and $82$ parameters in Nazir's ansatz.

For each set of the coefficients ($\alpha, \beta, s$ and $\bar{s}$) in the continuous spectral densities $J_x(\omega)$ and $J_z(\omega)$, more than $100$ initial states are used in the iteration procedure with variational parameters ($A_n, B_n$) uniformly distributed within an interval $[-1, 1]$. Displacement coefficients $(f_{n,l}, g_{n,l})$ of the initial states obey the classical displacements, i.e., $f_{n,l} = -g_{n,l} \sim \lambda_{l} / 2\omega_{l}$ for the diagonal coupling bath and $f_{n,l} = -g_{n,l} \sim \phi_l/2\omega_{l}$ for the off-diagonal coupling bath. In the single-mode case, $f_{n,l}, g_{n,l}$, $A_n$  and $B_n$ are all initialized randomly. After preparing the initial state, a relaxation iteration technique \cite{zhao92,zhao95} is adopted, and a simulated annealing algorithm \cite{zhou14} is also employed to improve the energy minimization procedure. The iterative procedure is carried out until the target precision of $1\times10^{-12}$ is reached.

Theoretically, the number of coherent superposition states $N \rightarrow \infty$ is required for the completeness of the environmental wave function in variational method.
However, large values of $N$ pose significant challenges in carrying out numerical simulations. To obtain reliable numerical results with large $N$,
an approach to improve the variational algorithm is undertaken based on the parity symmetry. Assuming $|\Psi_g\rangle$ is the ground state obtained by the variational method with $N$ coherent superposition states, an intermediate state $|\Psi_{\rm int}\rangle$ can be generated via the parity symmetry operators ${\cal I},{\cal P}_{x},{\cal P}_{z}$ and ${\cal P}_x{\cal P}_z$,
\begin{equation}
|\Psi_{\rm int}\rangle = \left( C_1{\cal I}+C_2{\cal P}_{x}+C_3{\cal P}_{z}+C_4{\cal P}_x{\cal P}_z \right)|\Psi_g\rangle,
\label{intermediate}
\end{equation}
where $C_i(i=1,2,3,4)$ is the weight coefficient. According to the symmetry analysis \cite{zhou14}, the symmetry operator ${\cal P}_{x}$ or ${\cal P}_{z}$ can lead to the other branch of the doubly degenerate ground state with the same energy $E_{\rm g}$. Hence these four symmetry operators should be equally weighted with $|C_1|=|C_2|=|C_3|=|C_4|$. If $C_1=1$, $C_2=\pm 1, C_3=\pm 1$ and $C_3=\pm 1$ can be derived. Similar to the case of the delocalized Davydov $D_{\rm 1}$ variational ansatz in the Holstein model \cite{sun13}, the energy $E_{\rm int}$ of the intermediate state is lower than $E_{\rm g}^N$ after considering the parity symmetry. Using these eight states as initial states, one can obtain a new ground state $|\Psi_g\rangle$ by performing the variational procedure with $4N$ coherent superposition states, which yields a lower ground state energy $E_{\rm g}^{4N} < E_{\rm int} < E_{\rm g}^N$.

Due to numerical errors, however, the state $\Psi$ found by the variational algorithm corresponds only to the local minimum in energy in the vicinity of the ground state. To refine the variational results, the rotational symmetry should also be considered in the case of $s=\bar{s}$ and $\alpha = \beta$.  Via the rotational operator $\hat{T}(\theta)$ acting onto the state $|\Psi\rangle$, a subspace composed of a series of states with respect to the rotational angle $\theta$ is built.  Subsequently, the state with the minimum energy in this subspace is regarded as the ground state. Since the generater of the $U(1)$ symmetry $\hat{S}$ involves a hopping between the diagonal and off-diagonal coupling baths, the displacement coefficients in the two baths are identical after considering the rotational symmetry, consistent with the argument that the ground state is accompanied by a symmetric distribution of phonon numbers in the diagonal and off-diagonal coupling baths.

The phonon population $P_{\rm ph}(x,z)$ in Eq.~(\ref{density}) can be calculated with the multi-$D_{\rm 1}$ variational ansatz in Eq.~(\ref{vmwave}) as
\begin{equation}
P_{\rm ph}(x,z)=\sum_{n=1}^{N}\frac{\left[A_nF_{n}(x,z)\right]^2+\left[B_nG_{n}(x,z)\right]^2}{D},
\label{vm_density}
\end{equation}
where $D=\langle\Psi_g|\Psi_g\rangle$ is the norm of the wave function, $A_n$ and $B_n$ denote the weight coefficients of the $n$-th coherent superposition state coupled to the spin up and down states, and the phonon functions $F_n(x,z)=\langle \vec{r}|\psi_{+}\rangle_{\rm ph}=f_{n,x}(x)f_{n,z}(z)$ and $G_n(x,z)=\langle\vec{r}|\psi_{-}\rangle_{\rm ph}=g_{n,x}(x)g_{n,z}(z)$ represent the phonon component of the wave function $|\psi_{\pm}\rangle_{\rm ph}$ in the two-dimensional coordinate representation $\vec{r} = (x, z)$. The function $f_{n,x}(x)$ denoting a coherent state in the off-diagonal coupling bath can then be deduced,
\begin{eqnarray}
f_{n,x}(x) & = & \prod_l \langle x | f_{n,l} \rangle   \\
& = & \prod_l \left( \frac{\omega_l}{\pi}\right)^{1/4} e^{-ix_lp_l/2}e^{ip_lx}e^{-\omega(x-x_l)^2/2},  \nonumber
\end{eqnarray}
where $x_l$ and $p_l$ are defined as
\begin{eqnarray}
  p_l &  = & -i \sqrt{\frac{\omega_l}{2}}\left(f_{n,l} - f_{n,l}^{*}\right), \\
  x_l & = & \frac{1}{\sqrt{2\omega_l}}\left(f_{n,l} + f_{n,l}^{*}\right).
\end{eqnarray}
In the same way, the functions $f_{n,z}(z), g_{n,x}(x)$ and $g_{n,z}(z)$ can also be calculated with the displacement coefficients $f_{n,l}$ and $g_{n,l}$ in Eq.~(\ref{vmwave}) as input.
In the single-mode case, i.e., $M=1$, the phonon function can be simplified as $F_n(x,z)=f_{n,x}(x)f_{n,z}(z)=\langle z | f_{n,1} \rangle \langle x | f_{n,2} \rangle$ where the subscripts $1$ and $2$ correspond to the diagonal and off-diagonal coupling oscillators, respectively.

\section{single mode}

The ground state of the model involving two oscillators coupled diagonally and off-diagonally to a spin is investigated in this section. The corresponding Hamiltonian can be written as
\begin{eqnarray}\label{SBM_SB}
\hat{H}_{\textrm {single}} &=& \omega{(b_1^{\dagger}b_1+b_2^{\dagger}b_2)} +\frac{\sigma_z}{2}\lambda{(b_1^{\dagger}+b_1)} \\
&+& \frac{\sigma_x}{2}\phi{(b_2^{\dagger}+b_2)},
\end{eqnarray}
where $\lambda$ and $\phi$ are diagonal and off-diagonal coupling constants, respectively. It is the simplest version of the two-bath model under current study. Furthermore, we focus on the case of two identical coupling constants $\lambda=\phi$ as it gives the Hamiltonian the rotational symmetry, which may provide some simple insights on the nature of the phase transition between the critical and localized phases.

\subsection{Exact diagonalization}

In the exact diagonalization procedure, the phonon states $|\psi_{+}\rangle_{\rm ph}$ and $|\psi_{-}\rangle_{\rm ph}$ corresponding to the spin up and down states, respectively, are expanded in a series of Fock states,
\begin{equation}\label{Expand1}
|\psi_{+}\rangle_{\rm ph}=\sum_{k_1k_2}^{N_{\rm tr}}c_{k_1,k_2}|k_1k_2\rangle,
\end{equation}
\begin{equation}\label{Expand2}
|\psi_{-}\rangle_{\rm ph}=\sum_{k_1k_2}^{N_{\rm tr}}d_{k_1,k_2}|k_1k_2\rangle,
\end{equation}
where $c_{k_1,k_2}$ and $d_{k_1,k_2}$ are the coefficients of the Fock state $|k_1k_2\rangle$ for the two oscillators coupled diagonally and off-diagonally to the spin, and
$N_{\rm tr}=100$ is the bosonic truncation number defined as the cutoff value of the phonon occupation number. We have verified that this value of $N_{\rm tr}$ is sufficiently large for the ground-state energy to converge.
Solving the Schr\"{o}dinger equation, one can obtain the wave function of the ground state $|\Psi_g\rangle$ with a series of coefficients, $c_{k_1,k_2}$ and $d_{k_1,k_2}$, and the ground state energy $E_{\rm g}$. Thus, the expectation values of $\sigma_z$ and $\sigma_x$ can be calculated as
\begin{eqnarray}\label{Expectation}
\langle{\sigma_x}\rangle &=&\sum_{k_1k_2}^{N_{\rm tr}}c_{k_1k_2}^{*}d_{k_1k_2}+d_{k_1k_2}^{*}c_{k_1k_2}, \nonumber \\
\langle{\sigma_z}\rangle &=&\sum_{k_1k_2}^{N_{\rm tr}}|c_{k_1k_2}|^2-|d_{k_1k_2}|^2.
\end{eqnarray}
The phonon population $P_{\rm ph}(x,z)$ can also be obtained by
\begin{eqnarray} \label{ed_density}
&& P_{\rm ph}(x,z) =           \\ \nonumber
&& \sum_{k_1k_2}^{N_{\rm tr}}\left( \left[c_{k_1,k_2}C_{k_1,k_2}(x,z)\right]^2 + \left[d_{k_1,k_2}D_{k_1,k_2}(x,z)\right]^2\right),
\end{eqnarray}
where $C_{k_1,k_2}(x,z)$ and $D_{k_1,k_2}(x,z)$ represent the Fock states $|k_1k_2\rangle$ in the coordinate representation $\vec{r}=(x,z)$,
corresponding to the spin up and down states, respectively.

\subsection{Analytical results}
The characteristics of the single-mode spin-boson model involving two oscillators can be investigated intuitively in coordinate representation
with the transformation $\hat{x}=\left(b_{1}+b_{1}^{\dagger}\right)/\sqrt{2\omega}$ and $\hat{z}=\left(b_{2}+b_{2}^{\dagger}\right)/\sqrt{2\omega}$.
In the following discussion, we use $x$ and $z$  as the classical counterparts of the corresponding operators $\hat{x}$ and $\hat{z}$. The Hamiltonian is then described by
\begin{eqnarray}
H & = & H_{0}+V,   \nonumber \\
H_{0} & = & -\frac{1}{2}\nabla^{2}+\frac{1}{2}\omega^{2}r^{2},\\
V & = & \lambda^{\prime}\vec{r}\cdot\vec{\sigma}, \nonumber
\end{eqnarray}
where we denote $\vec{r}=(x,z)$, $r=\sqrt{x^{2}+z^{2}}$, $\vec{\sigma}=\left(\sigma_{x},\sigma_{z}\right)$, $\lambda^{\prime}=\lambda\sqrt{\omega/2}$, and $\nabla^{2}$ is the two-dimensional Laplace operator written in polar coordinates.
This Hamiltonian describes a spin in a two-dimensional harmonic potential
with spin-orbital coupling $V$. When $\lambda\gg$$\omega$, the
spatial motion of the particle is too slow compared to the degree of freedom
of the spin. Therefore, it is justifiable to introduce the Born-Oppenheimer
approximation. The spatial motion is thus treated
classically, and the spin-part is solved by
\begin{eqnarray}
V\left|\eta_{\pm}\right\rangle  & = & \epsilon_{\pm}\left|\eta_{\pm}\right\rangle
\end{eqnarray}
with the adiabatic eigenstates
\begin{eqnarray}
\left|\eta_{+}\right\rangle  & = & \left[\begin{array}{c}
\cos\left(\frac{\pi}{4}-\frac{\theta}{2}\right)\\
\sin\left(\frac{\pi}{4}-\frac{\theta}{2}\right)
\end{array}\right],\nonumber \\
\left|\eta_{-}\right\rangle  & = & \left[\begin{array}{c}
\sin\left(\frac{\pi}{4}-\frac{\theta}{2}\right)\\
-\cos\left(\frac{\pi}{4}-\frac{\theta}{2}\right)
\end{array}\right], \end{eqnarray}
and the corresponding eigenvalues $\epsilon_{\pm}=\pm\lambda^{\prime}r$.
Here we define $\tan\theta=z/x$.

Then, the wave function of the system can be assumed,
\begin{equation}
\left|\Psi\right\rangle =\varphi_{+}\left(\vec{r}\right)\left|\eta_{+}\right\rangle +\varphi_{-}\left(\vec{r}\right)\left|\eta_{-}\right\rangle .\label{eq:Psi}
\end{equation}
Using the adiabatic eigenstate as the basis, the equations for the spatial part of the wave function are obtained with the stationary
Schr\"{o}dinger equation $H\left|\Psi\right\rangle =E\left|\Psi\right\rangle $,
\begin{eqnarray}
&&(-\frac{\nabla^{2}}{2}+\frac{1}{8r^{2}}+\frac{\omega^{2}r^{2}}{2}+\lambda^{\prime}r\hat{\sigma}_{z}+i\hat{V}^{\rm n.a.}\hat{\sigma}_{y}-\hat{E})\vec{\Psi}(\vec{r}) \nonumber \\
&&=0,
\end{eqnarray}
where $\hat{E}=\mbox{diag}\left\{E_{+},E_{-}\right\}$, $\vec{\Psi}(\vec{r})=(\varphi_{+}(\vec{r}),\varphi_{-}(\vec{r}))^\top$, $\hat{\sigma}_{z}$ and $\hat{\sigma}_{y}$ are the Pauli matrixes, and the non-adiabatic terms are given by the operator
\begin{eqnarray}
\hat{V}^{\rm n.a.} & = & \frac{1}{2r^{2}}\frac{\partial}{\partial\theta},\label{eq:V}
\end{eqnarray}
which are only in the angular direction, and can be neglected in further analysis.
Following the standard procedure of variables' separation method, the solution has the following form
\begin{equation}
\varphi_{\pm}\left(\vec{r}\right)=\sum_{m}c_{m}e^{im\theta}R_{\pm}\left(r,m\right),
\end{equation}
wherein the radial function $R_{\pm}\left(r,m\right)$ is determined by
the equation
\begin{eqnarray}
\left[-\frac{1}{2r}\frac{\partial}{\partial r}\left(r\frac{\partial}{\partial r}\right)+V_{\mbox{eff}}\left(r\right)\right]R_{\pm}\left(r,m\right) & = & E_{\pm}R_{\pm}\left(r,m\right) \nonumber \\ & &
\label{eq:E4}
\end{eqnarray}
with the effective potential
\begin{equation}
V_{\mbox{eff}}\left(r\right)=\frac{\omega^{2}}{2}r^{2}\pm\lambda r+\left(\frac{1}{8}+\frac{m^{2}}{2}\right)\frac{1}{r^{2}}.\label{eq:Ve}
\end{equation}
The effective potential contains a harmonic potential, a linear potential
and a centrifugal potential, and the angular quantum number is half-integer
$\left(m=\pm\frac{1}{2},\pm\frac{3}{2}\dots\right)$ due to the contribution
of the spin $1/2$ part. Thus, the ground state that
corresponds to $m = \pm 1/2$ is doubly degenerate. In the strong coupling regime ($\lambda\gg\omega$),
we can neglect the centrifugal potential, leading to the expectation of a ring-shaped ground
state with the radius $R\propto\lambda/\omega^{2}$.

\begin{figure}[tbp]
\centering
\includegraphics*[width=1.0\linewidth]{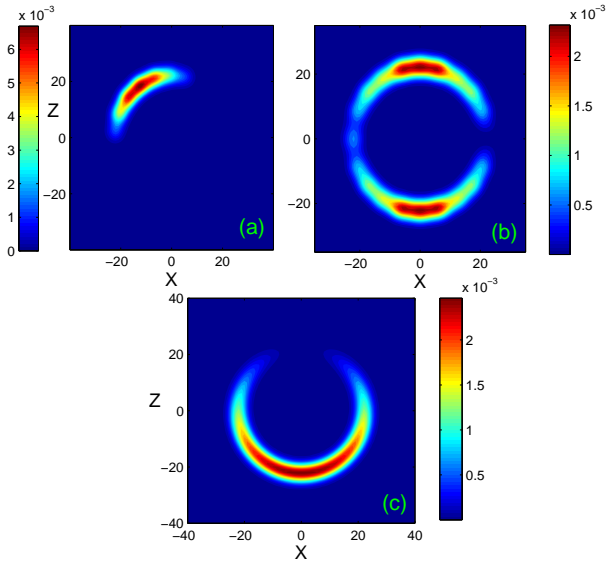}
\caption{ (Color on-line) The wave function of the ground state in a strong coupling case of $\lambda=\phi=10$ and $\omega=1$
is displayed in two-dimensional coordinate representation $(x,z)$.  The $x$- and $z$-coordinate
correspond to off-diagonal and diagonal coupling baths, respectively, and the colour represents the phonon population $P_{\rm ph}(x,z)$.
In (a) and (c), the numbers of coherent superposition states $N=8$ and $32$ are adopted, respectively, and an intermediate state defined in Eq.~(\ref{intermediate})
is shown in (b).
}
\label{f6}
\end{figure}

\subsection{Numerical results}

\begin{figure}[tbp]
\centering
\includegraphics*[width=1.2\linewidth]{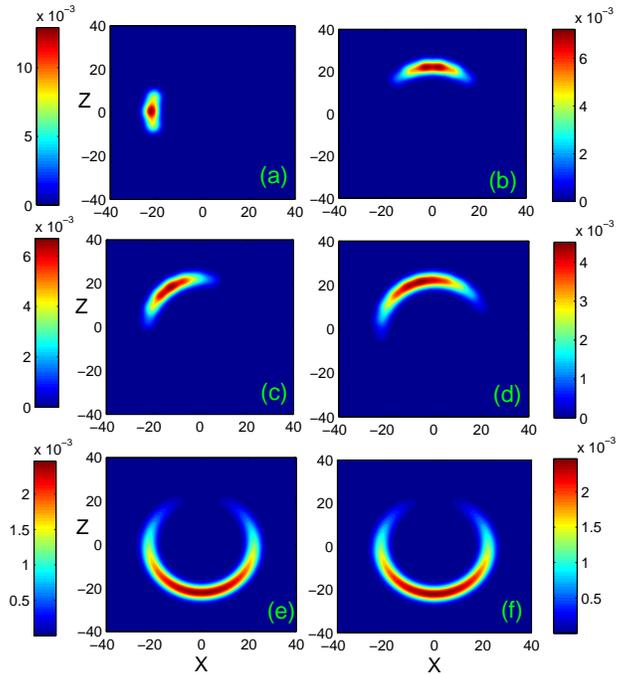}
\caption{ (Color on-line) The convergence test of the ground state at $\lambda=\phi=10$ and $\omega=1$
is displayed for various numbers of the coherent superpositions $N=4,6,8,12,32$ and $96$ in (a)-(f), respectively.
The $x$- and $z$-coordinate correspond to off-diagonal and diagonal coupling baths, respectively,
and the colour represents the phonon population $P_{\rm ph}(x,z)$. }
\label{f7}
\end{figure}

We first investigate the ground state of the two-bath model in the case of $\omega=1$ and $\lambda=\phi = 10$ via the variational method with $N=8$ and $M=1$.
According to the aforementioned theoretical arguments, the wave function of the ground state is expected to be ring shaped in the two-dimensional $(x,z)$ coordinate representation.
However, variational results depict only a quarter of the ring as shown in Fig.~\ref{f6}(a), where the colour represents the value of the phonon population $P_{\rm ph}(x,z)$ defined in Eq.~(\ref{vm_density}), and $x$- and $z$-coordinates correspond to the off-diagonal and diagonal coupling baths, respectively. Using the parity symmetry operators onto the ground state, an intermediate state defined in Eq.~(\ref{intermediate}) is obtained and shown in Fig.~\ref{f6}(b). The energy of this intermediate state, $E_{\rm int}=-25.48911811$, is found to be slightly below the ground state energy $E_{\rm g}^{N=8}=-25.48058088$. Taking this intermediate state as an initial state, one can seek the ground state via the variational method with $N=32$. Figure~\ref{f6}(c) shows the phonon population $P_{\rm ph}(x,z)$ decreasing smoothly with the $z$-coordinate, quite different from that of the intermediate state.
It indicates that the parity symmetry in the $z$ direction is broken, resulting in doubly degenerate ground states with different values of $\langle \sigma_z\rangle$ but the same ground state energy $E_{\rm g}^{N=32} = -25.49741979$ that is much lower than both $E_{\rm g}^{N=8}$ and $E_{\rm int}$.

\begin{figure}[tbp]
\centering
\includegraphics*[width=1.12\linewidth]{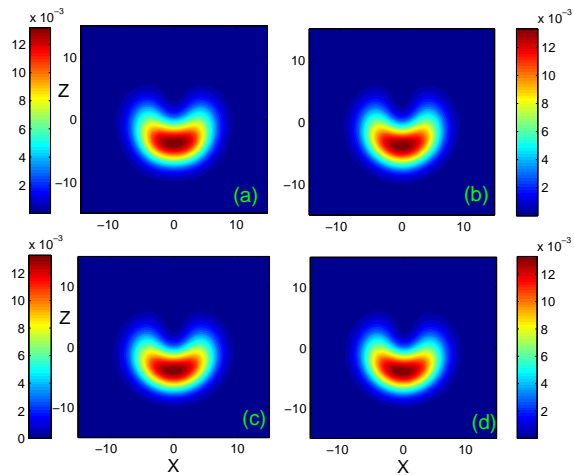}
\caption{ (Color on-line) The wave function of the ground state in a case of $\lambda=\phi=2$ and $\omega=1$
is displayed in (a)-(d) for various numbers of the coherent superposition states $N=4,6,16$ and $24$.
The $x$- and $z$-coordinate correspond to off-diagonal and diagonal coupling baths, respectively,
and the colour reflects the value of the phonon population $P_{\rm ph}(x,z)$. }
\label{f8}
\end{figure}

\begin{figure}[bp]
\centering
\includegraphics*[width=0.9\linewidth]{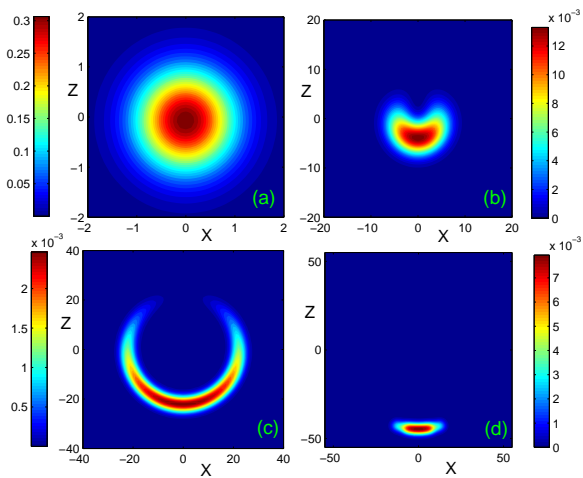}
\caption{ (Color on-line) The wave function of the ground state obtained by variational method
is displayed in (a)-(d) for the coupling strengths $\lambda=\phi=0.1,2,10$ and $20$, respectively.
The phonon frequency $\omega=1$ is set for both two baths. The $x$- and $z$-coordinate
correspond to off-diagonal and diagonal coupling baths, respectively,
and the colour represents the phonon population $P_{\rm ph}(x,z)$.}
\label{f10}
\end{figure}

Further, the convergence of ground state with respect to the number of the coherent superposition states $N$ warrants a careful examination. As shown in Fig.~\ref{f7}, the phonon population $P_{\rm ph}(x,z)$ gradually starts to resemble a ring-like shape as $N$ is increased from $4$ to $6, 8, 12, 32$ and $96$.
In fact, the $\rm X$-$\rm Z$ symmetric spin-boson model exhibits continuous degeneracy by the projecting $\hat{T}(\theta) |\Psi_g\rangle$, where $\hat{T}(\theta)$ is the rotational symmetry operator defined in Eqs.~(\ref{rotation}) and (\ref{generator}), and $|\Psi_g\rangle$ is one branch of the ground state. Moreover, the ground state energy $E_{\rm g}^N$ monotonically decreases with $N$ and is convergent to an asymptotic value $E_{\rm g}^{N=96}=-25.497421539$, consistent with the exact diagonalization result $E_{\rm g}=-25.497421544$ with the phonon truncation number $N_{\rm tr}=100$.

The ground state of the two-bath model in the case of weaker coupling, $\lambda=\phi=2$ and $\omega=1$, is investigated next. The phonon distribution $P_{\rm ph}(x,z)$ for $N=4,6,16$ and $24$ is displayed in Fig.~\ref{f8}(a)-(d). Different from the results shown in Fig.~\ref{f7}, the shape of the ground state remains nearly unchanged, indicating that a small value of $N$ is sufficient to obtain a reliable numerical result. The ground state energy $E_{\rm g}^{N=24}=-1.368929967$ is again in an excellent agreement with the exact diagonalization result $E_{\rm g} = -1.368929970$.

Finally, the ground states of two-bath model with $\lambda=\phi=0.1$ and $20$ are also plotted in Fig.~\ref{f10}(a) and (d), respectively, to facilitate comparison with those for $\lambda=\phi=2$ in Fig.~\ref{f10}(b) and $\lambda=\phi=10$ in Fig.~\ref{f10}(c). In the weak coupling regime $\lambda \ll \omega$, a ground state with a clear rotational symmetry is found, while it collapses in a corner of the $\rm X$-$\rm Z$ plane in the strong coupling regime $\lambda \gg \omega$. It supports our conjecture that the rotational symmetry breaks when the coupling strength exceeds a certain value $\lambda_c$, similar to the picture of the phase transition in the classical XY model. Since the radius of the circle in Fig.~\ref{f10}(a) is quite small, any slight shift of the center from the coordinate origin $(0,0)$ will induce a sharp jump in the spin polarization from $m=0$ to $m \approx \pm 1$. It indicates that the spin polarization $m$ is unstable in the weak coupling regime, corresponding to the free phase. That the ground state in Fig.~\ref{f10}(c) shows a crescent profile rather than a complete ring may be reasoned from previous analytical results. The ground state here must be doubly degenerate, and our numerical calculations yield only one branch of the ground state. Upon combining both the degenerate sates, once can readily obtain the complete ring shape of the ground state.

\begin{figure}[tbp]
\centering
\includegraphics*[width=0.9\linewidth]{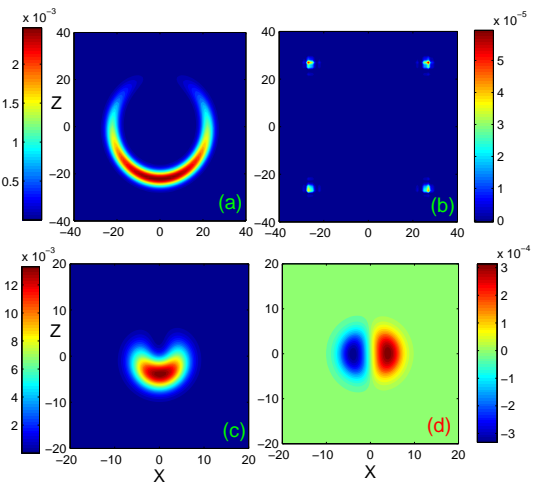}
\caption{ (Color on-line) The wave function of the ground state obtained by exact diagonalization
is displayed in (a) and (c) for $\lambda=\phi=10$ and $\lambda=\phi=2$, respectively.
Correspondingly, the difference between the exact diagonalization and variational results are displayed in (b) and (d).
The $x$- and $z$-coordinate correspond to off-diagonal and diagonal coupling baths, respectively,
and the colour represents the phonon population $P_{\rm ph}(x,z)$. }
\label{f9}
\end{figure}

\subsection{Discussion}

The ground sates obtained by the exact diagonalization method are shown in Figs.~\ref{f9}(a) and \ref{f9}(c) for the two cases of $\lambda=\phi=10$ and $\lambda=\phi=2$, respectively. Results for both of these cases seem to be identical to those with the variational results shown in Figs.~\ref{f7}(f) and \ref{f8}(d). To further verify the consistency, the difference between exact diagonalization and variational results are displayed in Figs.~\ref{f9}(b) and (d). The resulting difference is two orders or magnitude smaller than the phonon population $P_{\rm ph}(x, z)$, implying that the wave function obtained by the two methods are nearly the same, thereby lending support to the superior accuracy of our variational results.

\begin{table}[tbp]\centering
\caption{ The ground state energy $E_{\rm g}$ and spin polarization $m$ obtained by the variational method (VM) and exact diagonalization (ED) are listed for three different cases with the diagonal and off-diagonal coupling strengths $\lambda=\phi=0.1, 2$ and $10$, respectively. The phonon frequency $\omega=1$ is set, and $N$ and $N_{\rm tr}$ denote the coherent-superposition number in the variational method and the bosonic truncated number in the exact diagonalization, respectively.  }
\begin{tabular}[t]{l| c c c c}
\hline \hline
              &                   &   $\lambda=\phi=0.1$   &  $\lambda=\phi=2$    &    $\lambda=\phi=10$  \\
\hline
VM            &     $N$           &    $8$                 &  $24$           &   $96$                 \\
              &     $E_{\rm g}$         &  $-4.987582654$E$-3$  &  $-1.368929967$ &  $-25.497421539$       \\
              &     $m$           &  $~0.995049457657$     &  $~~0.568606150$ &  $~~0.500053226$        \\

              &                   &                        &                 &                         \\
\hline
ED            &     $N_{\rm tr}$  &  $100$                 &  $100$          &   $100$                     \\
              &     $E_{\rm g}$         &  $-4.987582654$E$-3$  &  $-1.368929970$ &  $-25.497421544$        \\
              &     $m$           &  $~0.995049457657$     &  $~~0.568606143$ &  $~~0.500053212$        \\

\hline \hline
\end{tabular}
\label{t1}
\end{table}

\begin{figure}[bp]
\centering
\includegraphics*[width=0.8\linewidth]{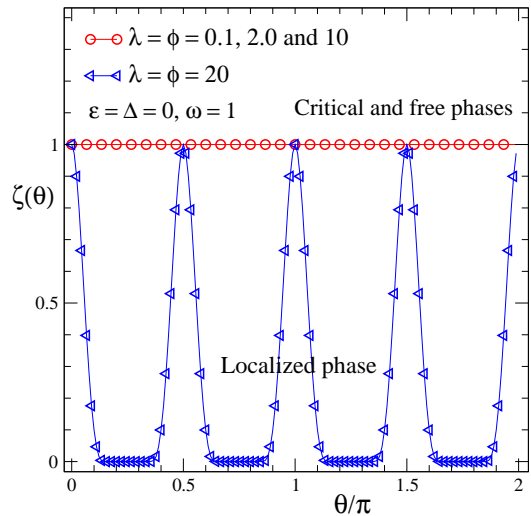}
\caption{ (Color on-line) The parity symmetry parameter $\zeta(\theta) = \sqrt{\zeta_x^2+\zeta_z^2}$
is displayed as a function of the rotation angle $\theta/\pi$ for coupling strengths $\lambda=\phi$.
The spin bias $\varepsilon = 0$, tunneling constant $\Delta=0$,
and frequency $\omega=1$ are set. The dash lines represent the fitting with the trigonometric functions.}
\label{f11}
\end{figure}

\begin{figure}[tbp]
\centering
\includegraphics*[width=0.8\linewidth]{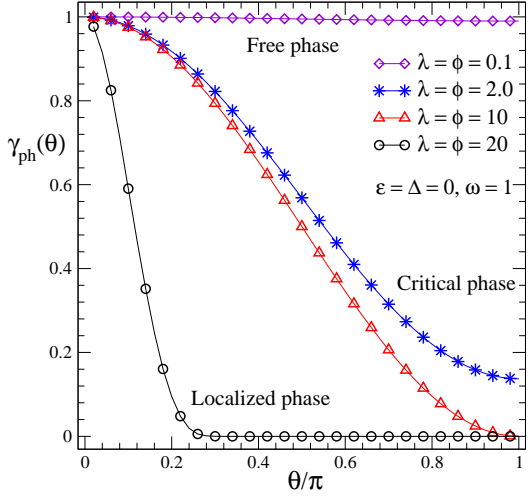}
\caption{ (Color on-line) The rotation symmetry parameter $\gamma_{\rm ph}(\theta)$ is displayed as a function of the rotation angle $\theta/\pi$
for various coupling strengths $\lambda=\phi=0.1, 2, 10$ and $20$. The spin bias $\varepsilon = 0$, tunneling constant $\Delta=0$,
and frequency $\omega=1$ are set.}
\label{f12}
\end{figure}

The ground state energy $E_{\rm g}$ and spin polarization $m$ obtained by the variational method and the exact diagonalization approach are summarized in Table.~\ref{t1} for the coupling strengths $\lambda=\phi=0.1,2$ and $10$. In all three cases, results from both the methods agree to each other for more than $9$ significant digits of $E_{\rm g}$ and $m$. Moreover, the radii of the rings $R=5, 25$ and $50$ in the cases of $\lambda=\phi=2, 10$ and $20$ calculated from Figs.~\ref{f8}(d), \ref{f7}(f) and \ref{f10}(d), respectively, are found to be consistent with the theoretical prediction $R = c\lambda/\omega^2$ with the coefficient $c=2.5$. This excellent reproduction of results again points to the superiority of the variational method and to the robustness of the ground state obtained by numerical calculations.

The symmetry of the ground state in the single-mode case is also studied via the symmetry parameters
$\zeta=\sqrt{\zeta_x^2+\zeta_z^2}$ of the parity symmetry and $\gamma_{\rm ph}$
of the rotational symmetry. Though the phase transitions may be reduced to the ground-state level crossings due to the finite number of degrees of freedom,
the symmetry properties of the ground states in the localized ($\lambda=\phi \gg 1$), critical, and free phases ($\lambda=\phi \ll 1$) are unchanged.
Fig.~\ref{f11} shows $\zeta(\theta)$ for $\lambda=\phi=0.1, 2, 10$ and $20$ when $\varepsilon=\Delta=0$ and $\omega=1$. Interestingly, it is found that $\zeta=1$ regardless of $\theta$ in the weak and intermediate coupling regimes, pointing to the parity symmetry in the critical and free phases.
In the localized phase, however, narrow peaks of $\zeta(\theta)$ are found for a strong coupling strength $\lambda=20$. It indicates that the ground state is localized without the parity symmetry. In Fig.~\ref{f12}, the symmetry parameter $\gamma_{\rm ph}(\theta)$ of the rotational symmetry is also plotted, which shows an abrupt decay to zero in the strong coupling regime ($\lambda=\phi=20$) but remains equal to one in the weak coupling regime ($\lambda=\phi=0.1$). In the intermediate regime, $\gamma_{\rm ph}(\theta)$ is found to decrease gradually. These numerical results further support our contention that the rotational symmetry breaks only when the coupling is strong.

\section{Continuous spectral densities}

\subsection{The case with $\alpha=\beta$}

\begin{figure}[tbp]
\centering
\includegraphics*[width=0.8\linewidth]{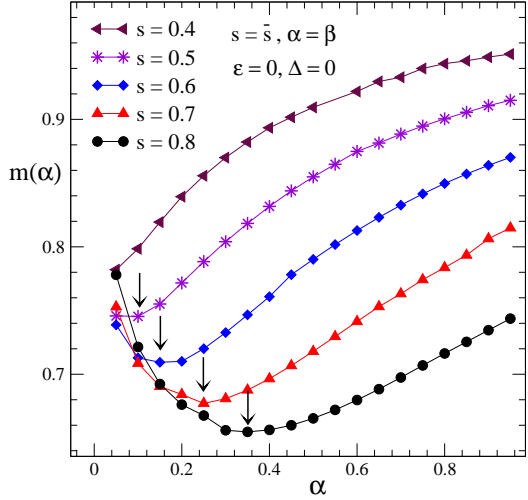}
\caption[FIG]{(Color online) The spin polarization $m(\alpha)$ at various values of the spectral exponent $s$ is plotted as a function of
the coupling strength $\alpha$ in the case with $\alpha=\beta$ and $s=\bar{s}$. The number of coherent superposition states $N=4$ and
effective bath modes $M=20$ are used in variational calculations. The downward arrows indicate the transition points $\alpha_{\rm c}$. }
\label{iden_1}
\end{figure}

In this subsection, we study the ground state properties of the two-bath model involving the baths described by a continuous spectral density function $J(\omega)$ via the variational approach. Infinite bath modes are considered in the variational calculations, although the number
of the effective modes $M$ is finite in the logarithmic discretization procedure. For convenience, we first examine the case involving two identical baths, i.e, $s=\bar{s}$ and $\alpha=\beta$.

The spin polarization $m$ defined in Eq.~(\ref{spin polarization}) is displayed in Fig.~\ref{iden_1} as a function of the coupling strength $\alpha$
for various values of the spectral exponent $s=0.4,0.5,0.6,0.7$ and $0.8$ in the case of $s=\bar{s}$ and $\alpha=\beta$.
For $\alpha > \alpha_{\rm c}$, an increase of the spin polarization $m(\alpha)$ is found for all of $s$, corresponding to the localized phase shown in Fig.~\ref{f0}(b).
However, a non-zero spin polarization is found in the critical phase with $\alpha < \alpha_{\rm c}$, quite different from the prediction of $m=0$ by an earlier study \cite{guo}. In addition, the localized-to-critical transition point $\alpha_{\rm c}$ marked by the downward arrows is shifted visibly with an increase in $s$ except for $s=0.4$ for which no phase transition occurs. It indicates the critical value of the spectral exponent is $s^{*}\approx 0.5$, consistent with the prediction $s^{*}=1/2$ of the mean-field analysis \cite{zhou14}, but much smaller than $s^{*}=0.75(1)$ \cite{guo}.

\begin{figure}[tbp]
\centering
\includegraphics*[width=0.8\linewidth]{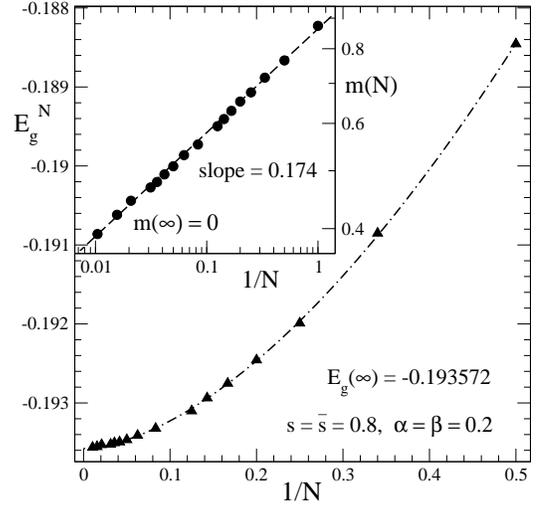}
\caption[FIG]{The ground state energy $E_{\rm g}^{N}$ is displayed as a function of the coherent-states number $N$ in a case of the critical phase with $s=\bar{s}=0.8$ and $\alpha=\beta=0.2$. The number of the effective bath modes $M=20$ is used in variational calculations. The dash-dotted line represents the fitting $E_{\rm g}^N= a N^{-b}+E_{\rm g}(\infty)$. In the inset, the spin polarization $m(N)$ is also shown on log-log scale, and the dashed line indicates a power law fit.}
\label{iden_2}
\end{figure}

\begin{figure}[bp]
\centering
\includegraphics*[width=0.8\linewidth]{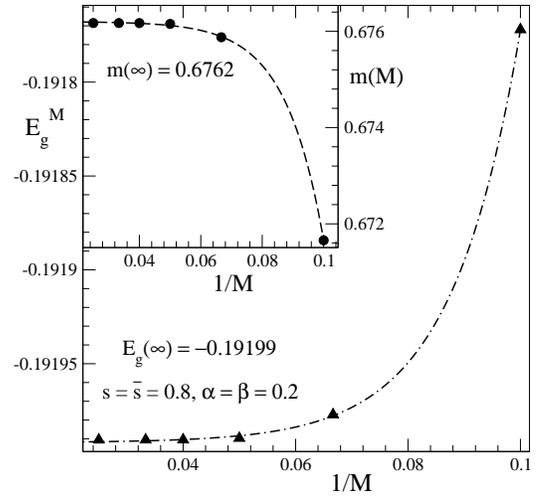}
\caption[FIG]{The ground state energy $E_{\rm g}^{M}$ and the spin polarization $m(M)$ are displayed as a function of the number of the effective bath modes $M$ at $N=4$ in a case of the critical phase case with $s=\bar{s}=0.8$ and $\alpha=\beta=0.2$. The dash-dotted and dashed lines represent the fitting $y(M) = a M^{-b}+ y(\infty)$. }
\label{iden_2_mod}
\end{figure}

\begin{table}[bp]\centering
\caption{ The ground state energy $E_{\rm g}$ and spin polarization $m$ obtained by the variational method (VM) and DMRG are listed for three different cases with ($s=\bar{s}=0.4, \alpha=\beta=0.1$), ($s=\bar{s}=0.6, \alpha=\beta=0.1$) and ($s=\bar{s}=0.8, \alpha=\beta=0.2$) in the localized phase (first case) and critical phase (last two cases). Three numbers of the coherent superposition states $N=16, 64$ and $96$ are used in variational calculations, which are sufficiently large in each case. $d_{\rm p}=50$ represents the phonon number allocated on each site on the Wilson chain in the DMRG algorithm. }
\begin{tabular}[t]{l| c c c c}
\hline \hline
              &                   &   $s=\bar{s}=0.4 \qquad$       &  $s=\bar{s}=0.6 \qquad$     &    $s=\bar{s}=0.8$  \\
              &                   &   $\alpha=\beta =0.1 \qquad$   &  $\alpha=\beta=0.1 \qquad$  &    $\alpha=\beta=0.2$                          \\
\hline
VM            &      $N$          &    $16$                &  $64$           &   $96$                 \\
              &     $E_{\rm g}$   &  $-0.16759$            &  $-0.12917$     &  $-0.19356$       \\
              &     $m$           &  $~0.77448$            &  $~0.53372$     &  $~0.40171$        \\

              &                   &                        &                 &                         \\
\hline
DMRG          &     $d_{\rm p}$   &  $50$                  &  $50$           &   $50$                     \\
              &     $E_{\rm g}$   &  $-0.16771$            &  $-0.12923$     &  $-0.19357$        \\
              &     $m$           &  $~0.75496$            &  $~0.42665$     &  $~0.29635$        \\

\hline \hline
\end{tabular}
\label{t2}
\end{table}

The convergence of the variational results with respect to $N$ and $M$ is carefully tested.
Fig.~\ref{iden_2} shows the ground-state energy $E_{\rm g}^N$ as a function of $N$ in the critical phase with $s=\bar{s}=0.8$ and $\alpha=\beta=0.2$. A power law decay of the ground state energy with the form $E_{\rm g}^{N}= a N^{-b}+E_{\rm g}(\infty)$ is found via numerical fitting, which yields the asymptotic value $E_{\rm g}(\infty)=-0.193572$.
In the inset, the spin polarization $m(N)$ is also displayed as a function of $N$ on a log-log scale. A perfect power-law behavior of $m(N)$ is obtained with the slope $0.174(2)$ and the asymptotic value $m(\infty)=0$. It suggests that the non-zero value of the spin polarization in the critical phase originates in the effects of the finite $N$. Furthermore, the spin polarization $m(N)$ in the critical phase is not convergent even for $N=96$, unlike the case of the localized phase where a small value of $N$ is sufficient to obtain reliable results. In a similar manner, the influence of $M$ to the ground state energy $E_{\rm g}^M$ and spin polarization $m(M)$ is also depicted in Fig.~\ref{iden_2_mod}. Both quantities are found to reach asymptotic values when $1/M < 0.05$, indicating the sufficiency of $M=20$. Therefore, in the following discussion on the variational results the number of coherent superposition states and the bath modes are set to $N=16$ and $M=20$, unless specified otherwise.

\begin{figure}[tbp]
\centering
\includegraphics*[width=0.8\linewidth]{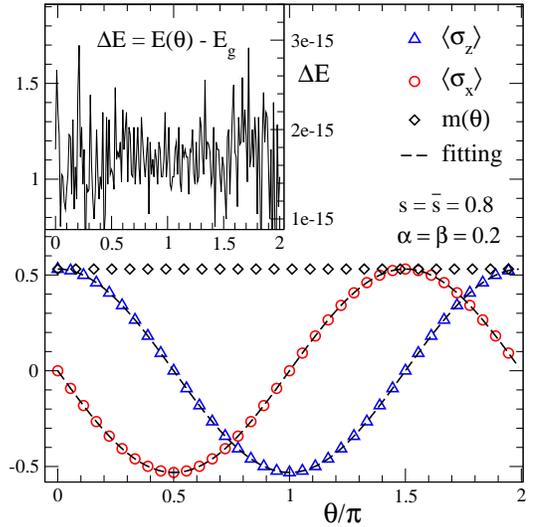}
\caption[FIG]{(Color on-line) The spin polarization $m$ and its $x$ and $z$ components $\langle \sigma_x\rangle$ and $\langle \sigma_z\rangle$ are plotted as a function of the
rotational angle $\theta$ for the states $\hat{T}(\theta)|\Psi_g\rangle$ in the case of $s=\bar{s}=0.8$ and $\alpha=\beta=0.2$. The dash lines represent the fitting with the trigonometric functions. In the inset, the shift $\Delta E=E(\theta) - E_{\rm g}$  from the ground state energy is shown. }
\label{iden_3}
\end{figure}

Table.~\ref{t2} presents a comparison between the numerical results obtained by the variational method and the DMRG approach \cite{Bulla,bulla2,Costi}. To ensure reliable results, the phonon number used in DMRG algorithm is $d_p=50$, much larger than $d_p=30$ used in the previous work \cite{guo}. The length of Wilson chain is set to $L=50$ and the cutoff dimension of the matrix is $D_c=60$. In the three cases of ($s=\bar{s}=0.4, \alpha=\beta=0.1$), ($s=\bar{s}=0.6, \alpha=\beta=0.1$) and ($s=\bar{s}=0.8, \alpha=\beta=0.2$), only a slight difference of the ground state energy, i.e., $\Delta E/E_g < 0.1\%$, is found between the variational and DMRG results, further reinforcing the superiority of our variational results. Since the ground state of the critical phase is unstable \cite{guo}, the spin polarization obtained by the variational method is larger than that by the DMRG, as shown in the last two columns in Table.~\ref{t2}. In the localized phase, however, a small value of $N=16$ is sufficient to obtain the variational result of the spin polarization $m=0.77448$, comparable with the DMRG result $m=0.75496$.

\begin{figure}[tbp]
\centering
\includegraphics*[width=0.8\linewidth]{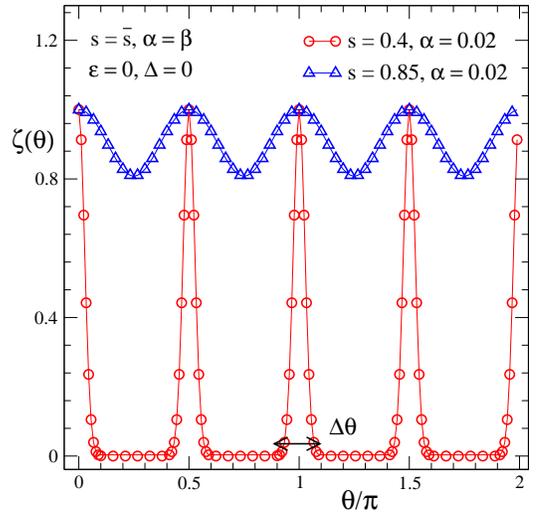}
\caption[FIG]{(Color on-line) The symmetry parameter of the parity symmetry  $\zeta(\theta)=\sqrt{\zeta_x(\theta)^2+\zeta_z(\theta)^2}$ is displayed as a function of $\theta/\pi$ for the two cases of $s=\bar{s}=0.4$ (bottom) and $s=\bar{s}=0.85$ (top) when the coupling strengths are $\alpha=\beta=0.02$. The width of the peak $\Delta \theta$ is defined as the size of the parity-symmetry regime with $\zeta >0$. }
\label{iden_4}
\end{figure}

Figure.~\ref{iden_3} shows the spin polarization $m(\theta)$ and its $x$ and $z$ components $\langle \sigma_x\rangle$ and $\langle \sigma_z\rangle$ for the states $\hat{T}(\theta)|\Psi_g\rangle$, where $\hat{T}(\theta)$ is the rotational symmetry operator defined in Eq.~(\ref{rotation}) and $|\Psi_g\rangle$ is the ground state obtained by the variational method. As the rotational angle $\theta$ increases, the values of $\langle \sigma_x\rangle$ and $\langle \sigma_z\rangle$ oscillate between $-0.5$ and $0.5$, while the corresponding spin polarization $m$ remains almost unchanged. The obtained curves can be fitted with trigonometric functions, $\langle \sigma_x\rangle = - m \sin(\theta)$ and $\langle \sigma_z\rangle = m \cos(\theta)$ with $m=0.53091$. It indicates that neither $\langle \sigma_x\rangle$ nor $\langle \sigma_z\rangle$ is a good candidate to characterize the localized-to-critical phase transition, even though they were employed in Ref.\cite{guo}. In the inset, the shift $\Delta E(\theta)=E(\theta)-E_{\rm g}$ is plotted. The sufficiently small value of $\Delta E \approx 2 \times 10^{-15}$ shows that there are continuous degenerate ground states which have the same energy $E_{\rm g}$, consistent with the prediction from the rotational symmetry analysis.

\begin{figure}[tbp]
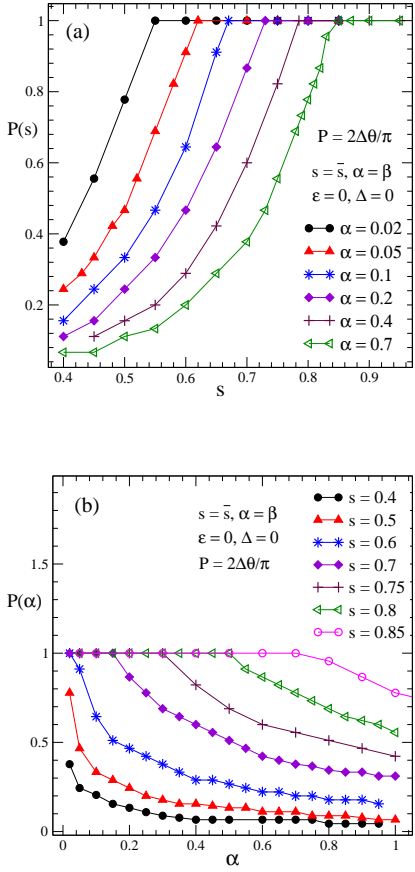

\centering
\begin{minipage}[b]{0.55\textwidth}
\hspace{-0.1\textwidth}
\includegraphics[width=0.55\linewidth]{xy_5.eps} \\
\vspace{2.7\baselineskip}
\hspace{-0.1\textwidth}
\includegraphics[width=0.55\linewidth]{xy_6.eps}
\end{minipage}
\caption[FIG]{ (Color on-line) The parity index, $P(\alpha, s)=2\Delta\theta/\pi$, is displayed in (a) as a function of $s$ for various values of $\alpha$ and in (b) as a function of $\alpha$ for various values of $s$. The transition point separating the localized phase from the critical phase is located at the position where the parity index $P=1$ reaches. }
\label{iden_5}
\end{figure}

Apart from the spin polarization, we have also probed the symmetry properties of the ground state.
The symmetry parameter of the parity symmetry, $\zeta(\theta)=\sqrt{\zeta_x(\theta)^2+\zeta_z(\theta)^2}$ where $\zeta_x$ and $\zeta_z$ are defined in Eq.~(\ref{order_parameter}), is displayed in Fig.~\ref{iden_4} for the two cases of $s=\bar{s}=0.4$ and $0.85$ at a sufficiently small coupling strength $\alpha=\beta=0.02$. In the localized phase with $s=0.4 < s^{*}$, sharp peaks of $\zeta(\theta)$ are found at $\theta/\pi \approx n/2$ ($n=1,2, 3$ and $4$) with a small peak width $\Delta \theta$ defined as the size of the parity-symmetry regime $\zeta>0$. It suggests that the ground state is localized in a corner of the $\rm X$-$\rm Z$ plane. On the other hand, $\zeta(\theta)$ is always greater than zero in the critical phase with $s=0.85 > s^{*}$, indicating that the parity symmetry covers the whole subspace. Hence, the parity index $P=2\Delta\theta/\pi$ reflecting the localization of the ground state can be used to quantitatively distinguish the localized and critical phases.

\begin{figure}[tbp]
\centering
\includegraphics*[width=0.8\linewidth]{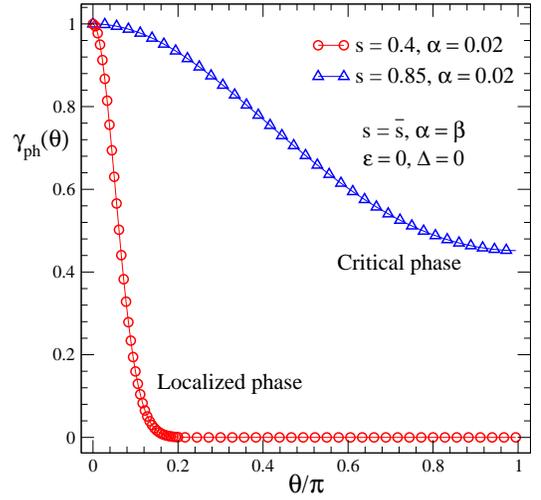}
\caption[FIG]{(Color on-line) The symmetry parameter of the rotation symmetry $\gamma_{\rm ph}(\theta)$ is displayed against the rotational angle $\theta/\pi$ for the
two cases of $s=\bar{s}=0.4, \alpha=\beta=0.02$ and $s=\bar{s}=0.85, \alpha=\beta=0.02$, corresponding to the critical and localized phases, respectively. }
\label{iden_6}
\vspace{2.8\baselineskip}
\end{figure}

Shown in Fig.~\ref{iden_5}(a) is the parity index $P(\alpha, s)$ versus the spectral exponent $s$ for various values of $\alpha$, in the case with two identical baths, i.e., $s=\bar{s}$ and $\alpha=\beta$. Without any loss of generality, it is assumed that only the ground state with the parity index $P =1$ belongs to the critical phase, otherwise it belongs to the localized phase. According to this criteria, the transition point $s_{\rm c}$ between the localized and critical phases is calculated for various values of $\alpha$. With an increase in $\alpha$, $s_{\rm c}$  increases monotonically, in agreement with the trend shown in Fig.~\ref{f0}(b). Moreover, the parity index $P(\alpha, s)$ is also displayed in Fig.~\ref{iden_5}(b) for various values of $s$. The transition point $\alpha_{\rm c}$ can be measured as a function of $s$ in a similar manner and subsequently the phase boundary in the $\rm X$-$\rm Z$ plane can be obtained.

\begin{figure}[tbp]
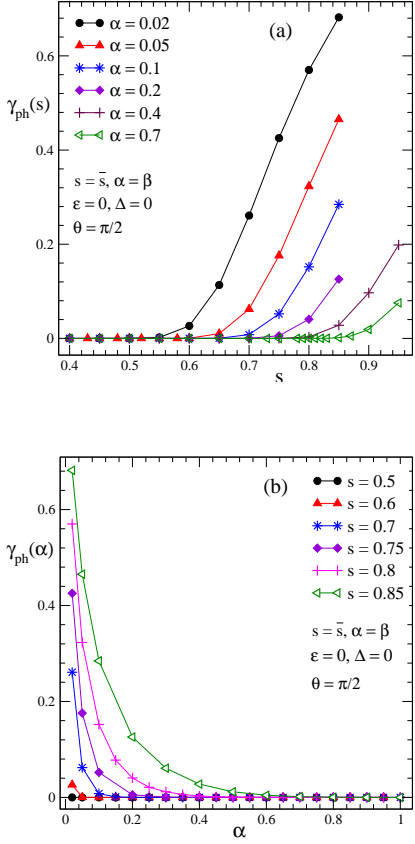

\centering
\begin{minipage}[b]{0.55\textwidth}
\hspace{-0.1\textwidth}
\includegraphics[width=0.55\linewidth]{xy_14.eps} \\
\vspace{2.5\baselineskip}
\hspace{-0.1\textwidth}
\includegraphics[width=0.55\linewidth]{xy_15.eps}
\end{minipage}
\caption[FIG]{ (Color on-line) The symmetry parameter of the rotational symmetry $\gamma_{\rm ph}(\alpha, s)$ at the angle $\theta=1/2\pi$ is displayed in (a) as a function of  $s$ for various values of $\alpha$ and in (b) as a function of $\alpha$ for various value of $s$. The transition point is located at the position separating the localized phase with $\gamma_{\rm ph} = 0$ from the critical phase with $\gamma_{\rm ph} > 0$. }
\label{iden_7}
\end{figure}

The rotational symmetry is studied next in the critical and localized phases, using the two typical cases
of $s=\bar{s}=0.4, \alpha=\beta=0.02$ and $s=\bar{s}=0.85, \alpha=\beta=0.02$, respectively.
The symmetry parameter $\gamma_{\rm ph}(\theta)$ is displayed in Fig.~\ref{iden_6} by rotating the two baths in the $\rm X$-$\rm Z$ plane through an angle $\theta$. In the localized phase, $\gamma_{\rm ph}(\theta)$ quickly depletes to zero, different from that in the critical phase, where it gradually decays to a nonzero value. Compared to the results of the single-mode case in Fig.~\ref{f12}, one can find that the localized and critical phases correspond to the strong and intermediate coupling regimes, respectively. In general, the strong coupling regime has a large coupling strength $\alpha$ and a small spectral exponent $s$, while opposite trends ensue in the intermediate coupling regime. For the spectral exponent $s>1$, however, the system always resides in the weak coupling regime, corresponding to the free phase.

Similar to the case of the parity symmetry, the symmetry parameter of the rotational symmetry $\gamma_{\rm ph}(\alpha, s)$ can also be used to distinguish the localized and critical phases. Without loss of generality, we set a special rotational angle $\theta=\pi/2$ where the phonons in the diagonal and off-diagonal coupling baths are interchanged. Thus, the value of the symmetry parameter is expected to be $\gamma_{\rm ph}=0$ in the localized phase and $\gamma_{\rm ph}>0$ in the critical phase. Fig.~\ref{iden_7}(a) shows $\gamma_{\rm ph}(s)$ as a function of $s$ for various values of $\alpha$. The transition point $s_{\rm c}$ is then located at the position separating the localized phase from the critical phase. It monotonously increases with $\alpha$, consistent with the results in Fig.~\ref{iden_5}(a). Moreover, $\gamma_{\rm ph}(\alpha)$ is also plotted in Fig.~\ref{iden_7}(b) as a function of $\alpha$ for various values of $s$. The transition boundary $\alpha_{\rm c}(s)$ separating the localized and critical phases can then be appropriately calculated.

\begin{figure}[tbp]
\centering
\includegraphics*[width=0.7\linewidth]{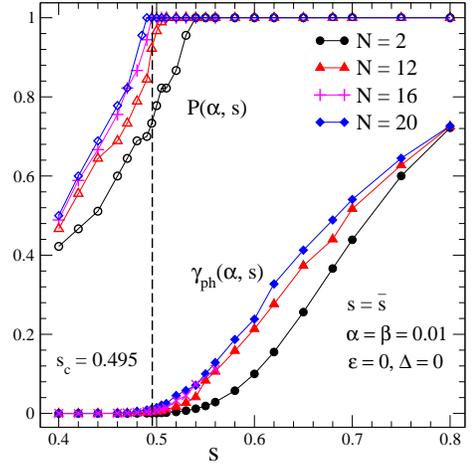}
\caption{ The rotation symmetry parameter $\gamma_{\rm ph}(\alpha, s)$ and the parity index $P(\alpha, s)$
for $N=2, 12, 16$ and $20$ are displayed as a function of $s$ in a weak coupling case of $\alpha=\beta = 0.01$. The dash line indicates the transition point $s_c=0.495(5)$.}
\label{iden_9}
\end{figure}

\begin{figure}[bp]
\centering
\includegraphics*[width=0.75\linewidth]{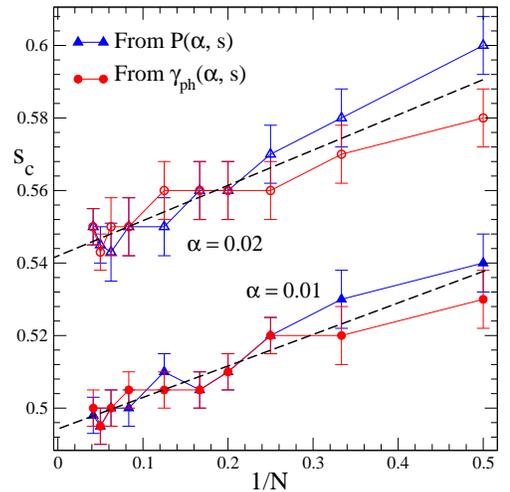}
\caption{Displayed as a function of the superposition number $N$ is the transition points determined by the variational calculations at weak coupling $\alpha=\beta=0.01$ (lower curves) and $0.02$ (upper curves). The circles and triangles correspond to the results from the curves of $\gamma_{\rm ph}(\alpha, s)$ and $P(\alpha, s)$, respectively. The dash lines are the linear fitting for the extrapolation of $s_c$ with $1/N \rightarrow 0$.}
\label{iden_95}
\end{figure}

To accurately estimate the critical value of the spectral exponent $s^{*}$ shown in Fig.~\ref{f0}(b), the case of very weak coupling strength $\alpha = \beta = 0.01$ is used to
investigate $\gamma_{\rm ph}(\alpha, s)$ and $P(\alpha, s)$.
As shown in Fig.~\ref{iden_9}, transition behavior of $P(\alpha, s)$ and $\gamma_{\rm ph}(\alpha, s)$
is displayed for $N=2, 12, 16$ and $20$.  The near overlap of the two curves of $N=16$ and $20$
suggests that $N=16$ is sufficiently large to accurately describe the phase transition.
With the increase of $N$, the transition point on the $P(\alpha, s)$ and $\gamma_{\rm ph}(\alpha, s)$ lines is found to decrease monotonically tending to an asymptotic value of $s_{\rm c}=0.495(5)$ as marked by the dashed line.
It points to the critical value of $s^* = 0.49(1)$ in the weak coupling limit of $\alpha \rightarrow 0$,
in perfect agreement with the mean-field prediction of $1/2$, but stands at variance with the value of $0.75(1)$ by Guo {\it et al.} \cite{guo}. This discrepancy may be attributed to the fact that the numerically unstable critical phase is beyond the reach of the DMRG algorithm of Guo {\it et al.}, and the external field holds great sway over the phase-transition properties of the two-bath model.

\begin{figure}[tbp]
\centering
\includegraphics*[width=0.75\linewidth]{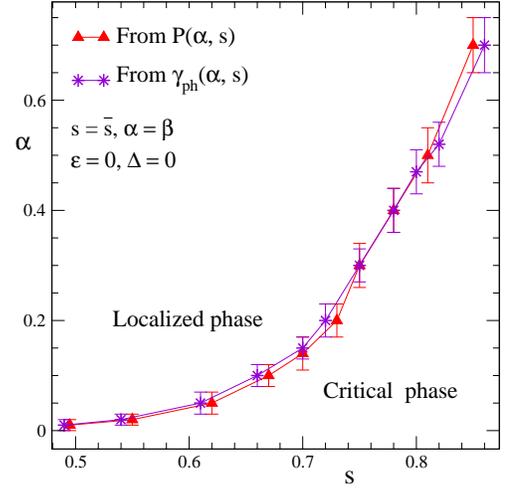}
\caption{ (Color on-line) The phase diagram of two-bath model is displayed in the $\alpha$-$s$ plane
in the case of $s=\bar{s}$ and $\alpha=\beta$. The phase boundary separating the critical phase from the localized phase is obtained from
the parity index $P(\alpha, s)$ and the symmetry parameter of the rotational symmetry $\gamma_{\rm ph}(\alpha, s)$.
}
\label{iden_8}
\end{figure}

In order to get a good estimate of $s_c(N)$, transition points calculated from the variational approach are in Fig.~\ref{iden_95} as a function of $1/N$ where $N$ is the number of superpositions.
Two values of coupling strengthes $\alpha=\beta=0.01$ and $0.02$ are used, and we also set $\varepsilon = \Delta=0$, and $s=\bar{s}$.
As $1/N$ decreases, the difference
between $\gamma_{\rm ph}(\alpha, s)$ and $P(\alpha, s)$ gradually disappears for both values of $\alpha$. Using linear fitting of $s_c(1/N)$, the asymptotic values of
$s_{\rm c}=0.493(6)$ and $0.541(7)$ for the two cases are obtained by extrapolation to infinite $N$, which is consistent with the $N=16$ results $s_c=0.500(5)$ and $0.55(1)$ within
the error bars, further supporting our claim that the number of superpositions $N=16$ is sufficient to obtain reliable results. The deviation of the critical point from the
Guo's result $s^*=0.75(1)$ is not induced by the effect of the finite value of $N$.
The phase diagram of the extended spin-boson model is displayed in Fig.~\ref{iden_8} in the case of $s=\bar{s}$ and $\alpha=\beta$. The solid triangles and stars represent the phase boundary obtained from the parity index $P(\alpha, s)$ and symmetry parameter of the rotational symmetry $\gamma_{\rm ph}(\alpha, s)$, respectively. The error bars in this phase diagram are estimated via the difference of the transition points measured with the fixed $\alpha$ and $s$, respectively.

\begin{figure}[bp]
\centering
\includegraphics*[width=0.75\linewidth]{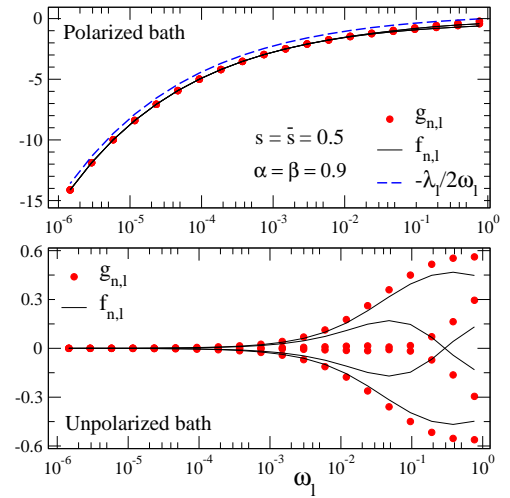}
\caption{(Color on-line) The displacement coefficients $f_{n,l}$ and $g_{n,l}$ of two-bath model in the localized phase are plotted as a function of the phonon frequency $\omega_l$
at $s=\bar{s}=0.5$ and $\alpha=\beta=0.9$. The upper and lower panels correspond to the polarized and unpolarized baths, respectively.
The dashed line represents the classical displacements rescaled by a factor $c=0.91$ for comparison. }
\label{ground_1}
\end{figure}

Having studied the symmetry properties in detail, we now turn our attention to the wave function of the ground state for the two-bath model involving the continuous spectral density to understand the nature of the localized and critical phases. To serve this purpose, we chose the case of $s=0.5, \alpha=0.9$ in the localized phase and the case of $s=0.85, \alpha=0.02$ in the critical phase as examples. Fig.~\ref{ground_1} shows the displacement coefficients $f_{n,l}$ and $g_{n,l}$ defined in Eq.~(\ref{vmwave}) in the localized phase as a function of the phonon frequency $\omega_l$. Two different behaviors of the displacement are found in the upper and lower panels for the phonons in the diagonal and off-diagonal coupling baths, corresponding to the ``polarized bath" and ``unpolarized bath," respectively. In the low frequency regimes, all the displacement coefficients $f_{n,l}$ and $g_{n,l}$ converge to a value independent of $n$, i.e., $f_{n,l}=g_{n,l}\rightarrow c \lambda_{l} / 2\omega_{l}$ in the polarized bath and $0$ in the unpolarized bath, where $c=-0.91$ is a $\omega$-independent constant.
In the low frequency regime, however, $f_{n,l}$ and $g_{n,l}$ exhibit quite different behaviors, and the relations $f_{1,l}=-f_{4,l}, f_{2,l}=-f_{3,l}$ and $g_{1,l}=-g_{2,l}, g_{2,l}=-g_{3,l}$ are found, indicating that the phonons in the unpolarized bath obey some kind of symmetry constraints. Moreover, the quantum fluctuations of the two-bath model in the localized phase are negligible, since the amplitude of the high-frequency displacements $A_p \approx 0.6$ in the unpolarized bath is much smaller than that of the classical displacement $|\lambda_{l} / 2\omega_{l}| \approx 15$ in the low-frequency limit.

\begin{figure}[tbp]
\centering
\includegraphics*[width=0.75\linewidth]{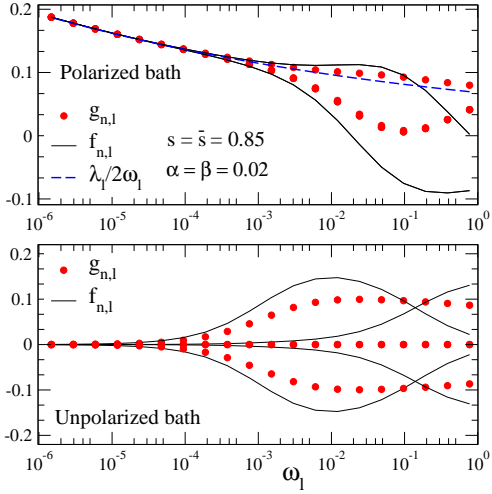}
\vspace{2\baselineskip}
\caption{(Color on-line) In the upper and lower panels, the displacement coefficients $f_{n,l}$ and $g_{n,l}$ in the critical phase
are plotted at $s=\bar{s}=0.85, \alpha=\beta=0.02$ for the polarized and unpolarized baths, respectively.
The dashed line represents the classical displacements rescaled by a factor $0.87$ for comparison. }
\label{ground_2}
\end{figure}

Fig.~\ref{ground_2} shows the displacement coefficients $f_{n,l}$ and $g_{n,l}$ in the the critical phase as functions of $\omega_l$. Similar to the results in the localized phase shown in Fig.~\ref{ground_1}, one bath of the model is in the polarized state, and the other in the unpolarized state. However, the classical displacement $f_{n,l} = g_{n,l} \rightarrow \lambda_{l} / 2\omega_{l} \approx 0.2 $ at $\omega_l = 10^{-6}$ in the upper panel is compatible with the amplitude of the high-frequency displacements $A_p \approx 0.15$ in the lower panel. It means that the quantum fluctuations play an important role in the critical phase, unlike the case in the localized phase.

\begin{figure}[tbp]
\centering
\includegraphics*[width=0.75\linewidth]{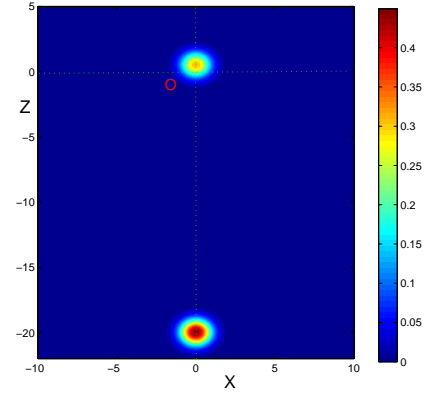}
\caption{(Color on-line) The wave function of the ground states is displayed for the two cases of $s=\bar{s}=0.85, \alpha=\beta=0.02$ (nearby the origin $O$)
and $s=\bar{s}=0.5, \alpha=\beta=0.9$ (in a corner), corresponding to the critical and localized phases, respectively.
The $x$- and $z$-coordinate correspond to off-diagonal and diagonal coupling baths, respectively,
and the colour represents the phonon population $P_{\rm ph}(x,z)$ at the frequency $\omega_l=10^{-6}$.
For convenience, we set the unit of the length in the $\rm X$-$\rm Z$ plane as $1/\sqrt{\omega_l}=10^3$.}
\label{ground_3}
\end{figure}

Finally, the ground states of the two-bath model in the localized and critical phases are compared via the phonon population $P_{\rm ph}(x,z)$ as shown in Fig.~\ref{ground_3}.
According to the results in Figs.~\ref{ground_1} and \ref{ground_2}, the displacement coefficients $f_{n,l}$ and $g_{n,l}$ in the localized and critical phases
are quite different in the low-frequency regime, especially at $\omega_l=10^{-6}$. Hence, only the bath modes at the frequency $\omega_l=10^{-6}$ are considered, and the unit of the length $1/\sqrt{\omega_l}=10^3$ is set. The phonon state in the case of $s=\bar{s}=0.85$ and $\alpha=\beta=0.02$ (critical phase) is located nearby the origin $O$, but the one in the case of $s=\bar{s}=0.5$ and $\alpha=\beta=0.9$ (localized phase) is far away it. In both the cases, the distance $d$ between the center of the phonon state
and the origin is proportional to the displacement $f_{n,l}, g_{n,l}$ in the polarized bath. Moreover, the central angle to the origin is calculated to be $\Theta=2\arctan(r/d)\approx 0.04\pi$ for the case of the localized phase, where $r=0.4$ is the radius of the phonon population $P_{\rm ph}(x,z)$ and $d=20$ is the distance.
Interestingly, the central angle $\Theta$ is comparable with the width of the peaks $\Delta\theta=0.038\pi$ defined in Fig.~\ref{iden_5}.
It further supports that the parity index $P=2\Delta \theta/\pi < 1$ reflects the localized nature of the ground state.

\subsection{The case with $\alpha\neq\beta$}
\begin{figure}[tbp]
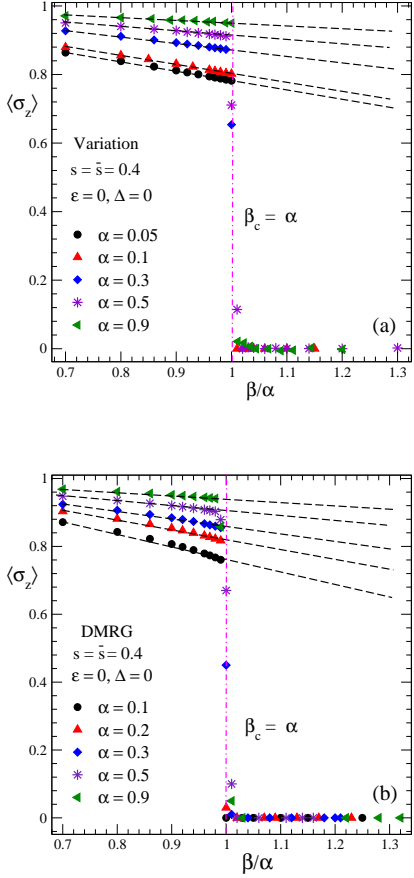

\centering
\begin{minipage}[b]{0.55\textwidth}
\hspace{-0.1\textwidth}
\includegraphics*[width=0.55\linewidth]{xy_9.eps} \\
\vspace{2.5\baselineskip}
\hspace{-0.1\textwidth}
\includegraphics*[width=0.55\linewidth]{DMRG_1.eps}
\end{minipage}
\caption[FIG]{(Color on-line)  The $z$ component of the spin polarization $\langle \sigma_z \rangle$ for various values of $\alpha$ is plotted as a function of
the ratio $\beta / \alpha$ in (a) and (b), corresponding to the variational and DMRG results, respectively. In both (a) and (b), the dash-dotted line indicates the transition point $\beta_c=\alpha$, and dashed lines represent linear fits. The value of the spectral exponent $s=\bar{s}=0.4$ is set.}
\label{multi_1}
\end{figure}

In the two bath model, we next investigate the case with $\beta \neq \alpha$ to further identify the critical and localized phases.
According to Ref.~\cite{zhou14}, there exists a first-order quantum phase transition separating the doubly degenerate ``localized state'' with $|\langle \sigma_z \rangle| >0$ and $\langle \sigma_x \rangle =0$ from the doubly degenerate ``delocalized state'' with $\langle \sigma_z \rangle =0$ and $|\langle \sigma_x \rangle| >0$. The transition point $\beta_c = \alpha$ is expected from the $\rm X$-$\rm Z$ symmetry when the spectral exponents obey $s=\bar{s}$.  In the following variational calculations, we use $N=4$ and $M=20$, which have been found to be sufficient in obtaining reliable results of the localized-to-delocalized phase transition \cite{zhou14}.

In Fig.~\ref{multi_1}(a), the $z$ component of the spin polarization $\langle \sigma_z \rangle$ obtained by the variational method is plotted with respect to the ratio $\beta / \alpha$ for various values of the diagonal coupling $\alpha$ in the case of $s=\bar{s}=0.4$. The transition point is determined at $\beta_c/\alpha = 1.0000(1)$, in perfect agreement with the expectation $\beta_c=\alpha$ within numerical errors. Furthermore, a linear decay of $\langle \sigma_z \rangle$ is found for $\beta < \beta_c$ before showing an abrupt jump to zero at the transition point, thereby verifying the transition to be of the first order. For comparison, the numerical results obtained by the DMRG algorithm are also displayed in Fig.~\ref{multi_1}(b). Similar behavior of $\langle \sigma_z \rangle$ is found, pointing to the validity of the variational method.

\begin{figure}[tbp]
\centering
\includegraphics*[width=0.7\linewidth]{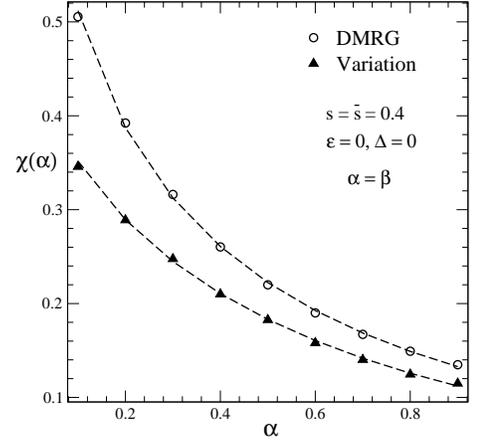}
\vspace{2\baselineskip}
\caption{ The generalized susceptibility $\chi=\left.\partial m / \partial \tau \right|_{\tau=0}$ is displayed as a function of the coupling strength $\alpha$
at $s=\bar{s}=0.4$. The circles and triangles correspond to the DMRG and variational results, respectively,
and the dashed lines represent exponential fits.
}
\label{multi_2}
\end{figure}

Due to infinite degenerate ground states $\hat{T}(\theta)|\Psi_g\rangle$, the spin polarization $\langle \sigma_z\rangle$ at the transition point $\beta_c=\alpha$ is variable as shown in Fig.~\ref{iden_3}.  In the regime with $\beta  < \alpha$, however, the value of $\langle \sigma_z \rangle$ is quite robust, since the rotational symmetry is broken by the anisotropic coupling. The generalized susceptibility $\chi$ is thus introduced by
\begin{equation}
\chi = \left.\frac{\partial m}{\partial \tau}\right|_{\tau=0},
\label{suscep}
\end{equation}
where $m$ is the spin polarization defined in Eq.~(\ref{spin polarization}), and $\tau=|\beta-\beta_c|/\beta_c= |\beta/\alpha - 1| $ is the reduced coupling strength. As shown in Fig.~\ref{multi_1}, the $z$ component of the spin polarization $\langle \sigma_z \rangle$ can be fitted by a linear behavior. Since the $x$ component of the spin polarization is
$\langle \sigma_x \rangle = 0$ in the localized state \cite{zhou14}, one obtains $m=\langle\sigma_z\rangle = a \tau + b$. The generalized  susceptibility $\chi=a$ can then be calculated with the linear fitting procedure for different values of $\alpha$ and $s$. An extended scaling form $m=1- a\exp(-b\tau)$ could lead to a better fitting of the numerical data, and yield $\chi=ab$ in the limit of $\tau \rightarrow 0$. Fig.~\ref{multi_2} shows the generalized susceptibility $\chi(\alpha)$ at $s=\bar{s}=0.4$ obtained by the variational method (solid triangles) and DMRG algorithm (open circles). Both of them decay exponentially with $\alpha$, though the exponents obtained via the fitting with $\chi \sim \exp(-c\alpha^d)$ differ for the DMRG ($d=0.5$) and variational method ($d=0.7$). The two curves $\chi(\alpha)$ follow the same behavior, implying that the system is always in the localized phase at $s=\bar{s}=0.4$ for various values of $\alpha$, and no phase transition occurs when $s=0.4 < s^{*}$.

\begin{figure}[tbp]
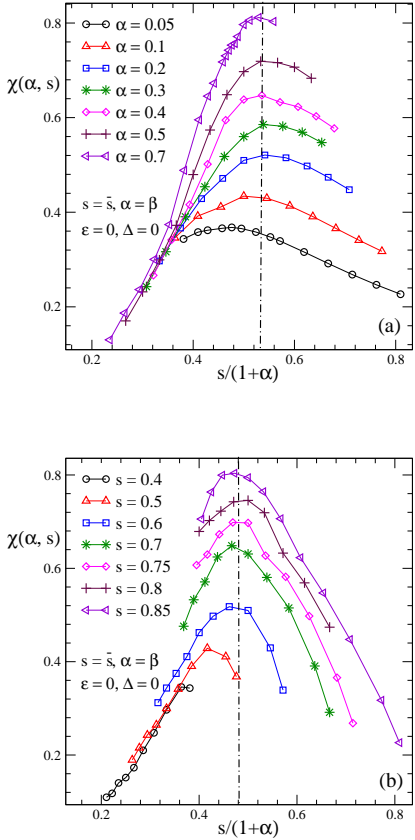

\centering
\begin{minipage}[b]{0.55\textwidth}
\hspace{-0.1\textwidth}
\includegraphics*[width=0.55\linewidth]{xy_10.eps} \\
\vspace{2.5\baselineskip}
\hspace{-0.1\textwidth}
\includegraphics*[width=0.55\linewidth]{xy_11.eps}
\end{minipage}
\caption{ (Color on-line) The generalized susceptibility $\chi(\alpha, s)$ is displayed as a function of $s/(1+\alpha)$ in (a)
for various values of $\alpha$ and in (b) for various values of $s=0.4$.
The dash-dotted lines indicate the peak positions of the curves for $\alpha > 0.1$ and $s>0.5$, pointing to a transition boundary $s/(1+\alpha) \approx 0.5$ in
the phase diagram.
} \label{multi_3}
\end{figure}

In Fig.~\ref{multi_3}(a), the generalized susceptibility $\chi(\alpha, s)$ is displayed as a function of the ratio $s/(1+\alpha)$ for various values of $\alpha$.
Unlike the trend shown in Fig.~\ref{multi_2}, $\chi(\alpha, s)$ increases with the spectral exponent
$s$ until $s/(1+\alpha) = 0.53$ marked by the dash-dotted line as the position of the peaks when $\alpha > 0.1$. It points to the transition boundary separating the localized phase from  the critical phase. Moreover, the generalized susceptibility $\chi(\alpha, s)$ is also displayed in Fig.~\ref{multi_3}(b) for various values of $s$. The position of the peak similarly marked by the dash-dotted line is found to be at $s/(1+\alpha)=0.48$ when $s > 0.5$. From these numerical results, one can obtain a relationship $s/(1+\alpha) \approx 0.5$, pointing to the transition boundary, i.e., $\alpha_{\rm c}=2(s-0.5)$. It further supports our contention that the critical value of the spectral exponent is $s^* \approx 0.5$.

\section{Conclusion}
The ground states of the spin-boson model with diagonal and off-diagonal coupling to two identical independent baths have been comprehensively studied in this paper by the variational approach, the DMRG algorithm and the exact diagonalization method. Adopting a generalized trial wave function, i.e., multi-${\rm D_1}$ ansatz, as the variational wave function, the spin polarization $m$, ground state energy $E_{\rm g}$ and wave function $|\Psi_g\rangle$ are calculated accurately by the variational method, in good agreement with those from the exact diagonalization in the case involving two oscillators and those from the DMRG algorithm in the case involving two baths describing a continuous spectral density function.

Three phases (localized, critical and free) are identified, corresponding to the strong, intermediate and weak coupling regimes, respectively.
Via the symmetry parameters $\zeta$ and $\gamma_{\rm ph}$,  the nature of the these phases is uncovered. The breakdown of the parity and rotational symmetries is found to occur along the quantum phase transition between the localized and critical phases. The phase boundary is determined by the parity index $P(\alpha, s)$ of the parity-symmetry regime with $\zeta > 0$, consistent with that obtained by $\gamma_{\rm ph}(\alpha, s)$. Moreover, the critical value of the spectral exponent is estimated as $s^*=0.49(1)$, well in agreement with the mean-field prediction $1/2$ \cite{zhou14}. The behavior of the spin polarization $m(\alpha, s)$ and the generalized susceptibility $\chi(\alpha, s)$ is also investigated for various values of the coupling strengthes $\alpha$ and spectral exponents $s$. Both the results point to $s^*\approx 0.5$, further supporting the accuracy of our results in pinning the transition point.

\section*{Acknowledgments}
The authors thank Bo Zheng, Yao Yao and Javier Prior for useful discussion.
Support from the Singapore National Research Foundation through the Competitive Research Programme
(CRP) under Project No.~NRF-CRP5-2009-04
is gratefully acknowledged. This work is also supported in part by and National Natural Science Foundation of China under Grant No.~$11205043$ and the U.S. National Science Foundation under Grant No. CHE-$1111350$.


%

\end{document}